\documentclass[traditabstract,final]{aa}
\usepackage{graphicx}
\usepackage[small]{caption}
\usepackage{a4wide}
\usepackage{epsfig}
\usepackage{natbib}
\usepackage{fancyhdr}
\usepackage{amsmath}
\usepackage{amssymb}
\usepackage{makeidx}
\usepackage{titletoc}
\usepackage[]{subfigure}
\usepackage{lscape}
\usepackage{hyperref}
\usepackage{wasysym}
\usepackage{wrapfig}
\usepackage{longtable}
\usepackage{aas_macros}
\usepackage{txfonts}
\bibpunct{(}{)}{;}{a}{}{,} % to follow the A&A style

\begin{document}

\title{Deep Wide-Field Imaging down to the oldest Main Sequence Turnoffs in the Sculptor dwarf spheroidal galaxy}
\titlerunning{Deep Wide-Field Imaging of  the Sculptor dSph}

   \author{T.J.L. de Boer\inst{1} \fnmsep \thanks{Visiting astronomer, Cerro Tololo Inter-American Observatory, National Optical Astronomy Observatory, which are operated by the Association of Universities for Research in Astronomy, under contract with the National Science Foundation.}
          \and
          E. Tolstoy\inst{1} \fnmsep \footnotemark[1]
          \and
          A. Saha\inst{2} \fnmsep \footnotemark[1]
          \and
          K. Olsen\inst{2} \fnmsep \footnotemark[1]
          \and
          M.J. Irwin\inst{3} 
          \and
          G. Battaglia\inst{4} 
          \and
          V. Hill\inst{5}
          \and
          M.D. Shetrone\inst{6}
          \and
          G. Fiorentino\inst{1} 
          \and
          A. Cole\inst{7} 
   } 
   \offprints{T.J.L. de Boer}

   \institute{Kapteyn Astronomical Institute, University of Groningen,  P.O. Box 800, 9700 AV Groningen, The Netherlands\\
              \email{deboer@astro.rug.nl}
             \and
              National Optical Astronomy Observatory \thanks{The National Optical Astronomy Observatory is operated by AURA, Inc., under cooperative agreement with the National Science Foundation.},
              P.O. box 26732, Tucson, AZ 85726, USA
             \and
              Institute of Astronomy, University of Cambridge, Madingley Road, Cambridge, UK, CB3 0HA
             \and
              European Southern Observatory, Garching bei Munchen, Germany 
             \and
              Observatoire de la C\^{o}te d'azur, CNRS UMR6202, Bd de l'Observatoire, BP 4229, 06304 Nice Cedex 4, France 
             \and
              McDonald Observatory, University of Texas at Austin, Fort Davis, TX
             \and
              School of Mathematics \& Physics , University of Tasmania, Hobart, Tasmania, Australia
             }

   \date{Received ...; accepted ...}
   
\abstract{We present wide-field photometry of resolved stars in the nearby Sculptor dwarf spheroidal galaxy using CTIO/MOSAIC, going down to the oldest Main Sequence Turn-Off. The accurately flux calibrated wide fieldÊColour-Magnitude Diagrams can be used to constrain the ages of different stellar populations, and also their spatial distribution. The Sculptor dSph contains a predominantly ancient stellar population ($>$10 Gyr old) which can be easily resolved into individual stars. A galaxy dominated by an old population provides a clear view of ancient processes of galaxy formation unimpeded by overlying younger populations. By using spectroscopic metallicities of RGB stars in combination with our deep Main Sequence Turn-Off photometry we can constrain the ages of different stellar populations with particular accuracy. 
We find that the known metallicity gradient in Sculptor is well matched to an age gradient. This is the first time that this link with age has been directly quantified. This gradient has been previously observed as a variation in Horizontal Branch properties and is now confirmed to exist for Main Sequence Turn-Offs as well. It is likely the Sculptor dSph first formed an extended metal-poor population at the oldest times, and subsequent more metal-rich, younger stars were formed more towards the centre until the gas was depleted or lost roughly 7 Gyr ago. The fact that these clear radial gradients have been preserved up to the present day is consistent with the apparent lack of signs of recent tidal interactions. }


\maketitle

\section{Introduction}
The Sculptor dwarf spheroidal galaxy is a faint~(M$_{V}$$\approx$$-$11.2), well studied system in the Local Group. It has a tidal radius of 76.5 arcmin on the sky~\citep{Irwin95}, which corresponds to 1.9 kpc at a distance of~86$\pm$5 kpc~\citep{Pietrzynski08}. It is located at high galactic latitude~(b$ = $-$83^{\circ}$) with a systemic velocity of V$_{hel} =+110.6 \pm 0.5$ km/s \citep{Battaglia082} and suffers from relatively low amounts of reddening, E(B$-$V)=0.018~\citep{Schlegel98}. Sculptor was the first early-type dwarf galaxy discovered around the Milky Way~\citep{Shapley38}. Since then it has been the target of numerous studies of its resolved stellar populations. \\
Early work on the radial and two-dimensional structure of the Sculptor dSph~\citep{Demers80,Eskridge881,Eskridge881,Irwin95} uncovered a complex spatial structure, with a radially increasing eccentricity. Furthermore, the presence of a spatial asymmetry was suggested in the East/West direction in observed star counts, of the order 30\% above the level expected for a symmetric profile~\citep{Eskridge882}. Using Colour-Magnitude Diagrams~(CMD), \citet{Norris78}, interpreting data from~\citet{Hodge65} and~\citet{Kunkel77}, argued that the Red Giant Branch~(RGB) is wider than can be explained by photometric uncertainties, suggesting the presence of internal age and/or abundance variations. The first CMDs using CCDs in a field just outside the core radius confirmed the large RGB spread and extended down to the Main Sequence Turn-Off~(MSTO) region, determining an age range of 13$\pm$2 Gyr~\citep{DaCosta84}. A very deep HST CMD of a small field of view~($\sim$2$^{\prime}$) well outside the centre of Sculptor, going down 3 magnitudes below the oldest MSTO, accurately confirmed the ancient age~(15$\pm$2 Gyr) of the bulk of the stars in Sculptor. This is similar to the age of globular clusters~\citep{Monkiewicz99}. From the same HST data set it was shown using CMD synthesis analysis that Sculptor displays an extended Star Formation History~\citep{Dolphin02}, with most of the stars having formed 8$-$15~Gyr ago, but with a small (and highly uncertain) tail reaching down to more recent times. Ground based wide-field imaging studies, covering a larger fraction ($\approx$30$^{\prime}$ $\times$ 30$^{\prime}$) of Sculptor, including the centre, found that the Horizontal Branch~(HB) morphology changes significantly with distance from the centre~\citep{Majewski99, HurleyKeller99}. \\
On the RGB the presence of the age-metallicity degeneracy is a problem for the accurate analysis of the properties of the Sculptor dSph. The age-metallicity degeneracy on the RGB results in two populations displaying nearly identical RGB, if the second population has an age a factor of 3 higher and a metallicity a factor of 2 lower than that of population 1. Therefore, spectroscopic observations are needed to remove the degeneracy on the RGB and allow an accurate analysis of Sculptor properties.  \\
Wide-field medium resolution \ion{Ca}{ii} triplet spectroscopy of RGB stars in Sculptor have independently confirmed the presence of two distinct stellar components~\citep{Tolstoy04, Coleman05, Westfall06}, with different spatial distributions, kinematics and metallicities. There is a spatially extended component which is more metal-poor ($-$2.8$<$[Fe/H]$<-$1.7), and a more concentrated, metal-rich component ($-$1.7$<$[Fe/H]$<-$0.9) which also has a lower velocity dispersion than the metal-poor component \citep{Battaglia07, Battaglia082}.  
\begin{figure}[!ht]
\centering
\includegraphics[angle=270, width=0.485\textwidth]{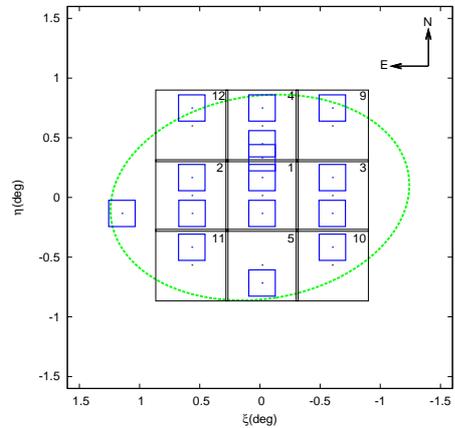}
\caption{Coverage of the photometric data across the Sculptor dwarf spheroidal. Black squares~(no. 1-5,9-12) denote the CTIO 4m fields observed. The 15 smaller blue squares show the CTIO 0.9m calibration fields that were observed. The green dashed ellipse is the tidal radius of Sculptor, as determined by~\citet{Irwin95}. \label{sclcov}} 
\end{figure}
\\
High resolution spectroscopic studies in Sculptor have been carried out to determine detailed abundances of a range of chemical elements for small numbers of individual RGB stars~\citetext{\citealp{Shetrone03} [5 stars]; \citealp{Geisler05} [4 stars]} using VLT/UVES and more recently with wide-field multi-fibre spectrograph VLT/FLAMES in high resolution mode (Hill et al., in prep; see~\citet{Tolstoy09}) for a sample of 93 stars. These studies are able to constrain the chemical evolution of Sculptor. This can then be compared to the other Local Group galaxies, including the Milky Way. For example, the [$\alpha$/Fe] abundances are sensitive to the relative rates of SNe II and SNe Ia and can thus be used to trace the star formation timescale. Thus, the [Fe/H] and [$\alpha$/Fe] abundances obtained from spectroscopic studies can be combined with deep photometry to accurately determine the ages of the Sculptor stellar populations, and the relation between Star Formation Rate and the chemical enrichment processes. Obtaining accurate time scales for star formation and the build up of abundance patterns in Sculptor also allows us to determine the early history of Sculptor, and also until which time Sculptor could contribute to the build-up of the Milky Way. \\
In this work we present deep wide-field carefully calibrated photometry of the Sculptor dSph galaxy, carried out with MOSAIC on the CTIO 4m/Blanco telescope. These CMDs reach down to the oldest MSTO, for an area covering $\approx$80\% of the tidal radius of Sculptor. Accurately calibrated photometry is essential in determining ages, since the position of the MSTO directly correlates with the age of the stellar population. From deep MSTO photometry over a large region we can for the first time study the effect of age on the physical properties of the galaxy. Combining the results of spectroscopic surveys with our MSTO photometry also allows us to measure the age-metallicity relation in Sculptor dSph galaxy. \\
The paper is structured as follows: in section~\ref{observations} we present the observations and data reduction. In section~\ref{photometry} we describe how we obtained and calibrated the photometry and the resulting structural properties. Section~\ref{interpretation} describes the analysis and interpretation of different evolutionary features in the CMDs. Finally, section~\ref{discussion} discusses the conclusions and their implications in terms of galaxy formation. In a subsequent paper we will present the detailed Star Formation History analysis of the Sculptor dSph.

\section{Observations \& Data Reduction}
\label{observations}
Deep optical photometry of the Sculptor dSph in the B, V and I bands was obtained using the CTIO 4-m MOSAIC II camera over 10 nights in September 2008 and November 2009. In order to ensure accurate photometric calibration without using too much 4m telescope time we obtained service mode observations with the 0.9m CTIO telescope over 3 photometric nights. Observations were made of Landolt standard fields~\citep{Landolt07, Landolt92} covering a range of different airmass and colour. \\
The positions of fields observed with the 4m~(big black squares) and 0.9m~(small blue squares) telescopes are shown in Fig.~\ref{sclcov}, relative to the Sculptor centre. B,V and I photometry was obtained for the central 5 pointings~(1,2,3,4,5), whereas the outer pointings~(9,10,11,12) were only observed in V and I bands. The observation logs for Sculptor observations are given in Tables~\ref{4mobs} and~\ref{0p9mobs}. The observation logs for standard star calibration fields with both the 4m and the 0.9m telescopes are given in Appendix~\ref{standardslist}.

\subsection{CTIO 4m MOSAIC}
The CTIO MOSAIC II camera has an array of eight 2048$\times$4096 pixel CCDs, each of which can be read out through one or two amplifiers.  The CCDs combine to form a 8192$\times$8192 pixel image corresponding to a field of view of 36$^{\prime}$ $\times$ 36$^{\prime}$ on the sky.  For our observing runs, one amplifier on one CCD was found to be inoperable, such that we were forced to read out MOSAIC II with 8 amplifiers only. This resulted in a readout overhead of 2m40s per image. An observing strategy was chosen involving non-dithered long~(600s) exposures to be stacked together. We chose not to dither images, in order to be able to directly co-add images to obtain the deepest photometry possible. Furthermore, not dithering allowed us to apply accurate position dependent aperture corrections for each pointing. The loss of area incurred by not dithering exposures is roughly 2 percent, which means it will not materially affect the conclusions in this paper. However, the very inner part of Sculptor~(r$_{ell}<$0.05 deg) does suffers from incomplete coverage due to the presence of CCD gaps there. Since the average saturation level of the CCDs is $\approx$43000 counts~(corresponding to B$\approx$18, V$\approx$17.5 and I$\approx$17.5 for 600s integration) additional shorter exposures~(90s + 10s) were taken to sample the bright stars in each field.  \\
For the data reduction we used the IRAF\footnote{IRAF is distributed by the National Optical Astronomy Observatories,which are operated by the Association of Universities for Research in Astronomy, Inc., under cooperative agreement with the National Science Foundation.} data reduction program, which includes the \textbf{MSCRED} package~\citep{Valdes98}, designed for working with mosaics of CCD's. The standard data reduction steps include correcting for bias and flatfield, removing bad pixels and obtaining precise astrometry. For wide-field images there are a number of extra steps which needed to be taken, as outlined in~\citep{Valdes02}. 
\begin{table*}[!ht]
\caption[]{List of fields observed in the Scl dSph with the 4m CTIO Blanco telescope. \label{4mobs}}
\footnotesize
\begin{center}
\begin{tabular}{lccccccc}
\hline\hline
Date 	& Field		&  RA		& DEC		& Filter	& exp time	& seeing	& airmass  \\
 	&				& J2000		& J2000		& 		& sec		& $^{\prime\prime}$ &        \\	
\hline
2009 Nov 19		   & Scl1	& 01:00:03.96	& $-$33:41:30.48		& B	& 1800, 90, 10		& 1.0-1.3 & 1.00-1.01	\\
2009 Nov 19 		   &		&		&			& V	& 1800, 90, 10		& 1.1-1.2 & 1.01-1.02	\\
2009 Nov 19 		   &		&		&			& I	& 2400, 90, 10		& 1.0-1.3 & 1.03-1.07	\\
2008 Sep 05 		   & Scl2	& 01:02:52.08 	& $-$33:41:24.00		& B	& 1500, 60		& 1.6-2.2 & 1.00-1.06	\\
2009 Nov 20		   &		&		&			& B	& 2400			& 1.6-2.2 & 1.00-1.06	\\
2008 Sep 05		   &		&		&			& V	& 1500, 60, 30		& 1.0-1.1 & 1.00-1.01	\\
2009 Nov 21		   &		&		&			& V	& 1800			& 1.0-1.1 & 1.00-1.01	\\
2008 Sep 05 		   &		&		&			& I	& 2400, 60,10		& 1.2-1.3 & 1.00-1.07	\\
2009 Nov 23 		   & Scl3	& 00:57:15.48 	& $-$33:41:22.92		& B	& 900, 60		& 1.2-1.4 & 1.62-1.70	\\
2009 Nov 23 		   &		&		&			& B	& 600			& 1.2-1.4 & 1.62-1.70	\\
2008 Sep 06 		   &		&		&			& V	& 900, 60, 30		& 1.0-1.2 & 1.01-1.02	\\
2009 Nov 21		   &		&		&			& V	& 1800			& 1.0-1.2 & 1.01-1.02	\\
2008 Sep 06		   &		&		&			& I	& 1800, 60, 10		& 0.9-1.0 & 1.03-1.08	\\
2009 Nov 21		   &		&		&			& I	& 2400			& 0.9-1.0 & 1.03-1.08	\\
2008 Sep 06 		   & Scl4	& 01:00:03.96 	& $-$33:06:30.60		& B	& 1500, 60, 10		& 0.9-1.0 & 1.01-1.03 	\\
2008 Sep 06		   &		&		&			& V	& 900, 60, 10		& 1.0-1.5 & 1.10-1.30	\\
2009 Nov 21 		   &		&		&			& V	& 1800			& 1.0-1.5 & 1.10-1.30	\\
2009 Nov 21 		   &		&		&			& I	& 2400, 90, 10		& 1.0-1.3 & 1.20-1.30	\\
2008 Sep 06 		   & Scl5	& 01:00:03.96 	& $-$34:16:30.36		& B	&  900, 60, 30		& 0.8-1.1 & 1.00-1.01	\\
2008 Sep 06 		   &		&		&			& V	& 900,60, 30		& 0.7-1.1 & 1.00-1.01	\\
2009 Nov 22		   &		&		&			& V	& 1800         		& 0.7-1.1 & 1.00-1.01	\\
2009 Nov 22 		   &		&		&			& I	& 2400, 90, 10		& 1.1-1.3 & 1.01-1.04	\\
2009 Nov 22 		   & Scl9	& 00:57:16.92 	& $-$33:06:23.40		& V	& 1800, 90, 10		& 1.1-1.2 & 1.06-1.11 	\\
2009 Nov 22 		   &		&		&			& I	& 2400, 90, 10		& 1.0-1.2 & 1.14-1.28	\\
2009 Nov 23 		   & Scl10	& 00:57:14.40 	& $-$34:16:22.80		& V	& 1800, 90, 10		& 1.1-1.2 & 1.10-1.17 	\\
2009 Nov 23 		   &		&		&			& I	& 2400, 90, 10		& 1.0-1.2 & 1.21-1.39	\\
2009 Nov 23 		   & Scl11	& 01:02:53.16 	& $-$34:16:23.52		& V	& 1800, 90, 10		& 0.9-1.0 & 1.00-1.02 	\\
2009 Nov 23 		   &		&		&			& I	& 2400, 90, 10		& 0.8-0.9 & 1.00-1.02	\\
2009 Nov 22 		   & Scl12	& 01:02:51.00 	& $-$33:06:24.12		& V	& 1800, 90, 10		& 1.0-1.1 & 1.29-1.43 	\\
2009 Nov 23 		   &		&		&			& I	& 2400, 90, 10		& 0.9-1.0 & 1.03-1.07	\\
\hline 
\end{tabular}
\end{center}
\end{table*}
\\
\begin{table*}[!ht]
\caption[]{List of fields observed in the Scl dSph with the 0.9m CTIO telescope. \label{0p9mobs}}
\footnotesize
\begin{center}
\begin{tabular}{cccccccc}
\hline\hline
Date	& Field		&  RA		& DEC		& Filter	& exp time	& seeing	& airmass \\
 	&	& J2000		& J2000		& 		& sec		& $^{\prime\prime}$ &           \\	
\hline
2008 Oct 31 & Scl1 N	& 01:00:03.96 	& $-$33:32:30.48	& B	& 300	& 1.5	& 1.02-1.05	\\
2008 Oct 31 &		&		&		& V	& 300	& 1.4	& 1.02-1.04 	\\
2008 Oct 31 &		&		&		& I	& 600	& 1.3	& 1.00-1.03	\\
2008 Nov 02 & Scl1 S	& 01:00:03.96 	& $-$33:50:30.48	& B	& 300	& 1.5	& 1.06-1.09	\\
2008 Nov 02 &		&		&		& V	& 300	& 1.3 	& 1.05-1.08	\\
2008 Nov 02 &		&		&		& I	& 600	& 1.3	& 1.04-1.07	\\
2008 Oct 31 & Scl2 N	& 01:02:52.08 	& $-$33:32:24.00	& B	& 300	& 1.7	& 1.01-1.03	\\
2008 Oct 31 &		&		&		& V	& 300	& 1.5	& 1.01-1.03	\\
2008 Oct 31 &		&		&		& I	& 600	& 1.5	& 1.02-1.04	\\
2008 Nov 02 & Scl2 S	& 01:02:52.08 	& $-$33:50:24.00	& B	& 300	& 1.3	& 1.02-1.04	\\
2008 Nov 02 & 		&		&		& V	& 300	& 1.3 	& 1.02-1.03	\\
2008 Nov 02 & 		&		&		& I	& 600	& 1.3	& 1.01-1.03	\\
2008 Oct 31 & Scl3 N	& 00:57:15.48 	& $-$33:32:22.92	& B	& 300	& 1.5	& 1.13-1.18	\\
2008 Oct 31 & 		&		&		& V	& 300	& 1.3 	& 1.15-1.20	\\
2008 Oct 31 & 		&		&		& I	& 600	& 1.3	& 1.16-1.22	\\
2008 Nov 02 & Scl3 S	& 00:57:15.48 	& $-$33:50:22.92	& B	& 300	& 1.8	& 1.01-1.03	\\
2008 Nov 02 & 		&		&		& V	& 300	& 2.0 	& 1.02-1.04	\\
2008 Nov 02 & 		&		&		& I	& 600	& 1.7	& 1.02-1.04	\\
2008 Oct 31 & Scl4 N	& 01:00:03.96 	& $-$32:57:30.60	& B	& 300 	& 1.5	& 1.25-1.35	\\
2008 Oct 31 & 		&		&		& V	& 300	& 1.3 	& 1.25-1.40	\\
2008 Oct 31 & 		&		&		& I	& 600	& 1.5	& 1.30-1.42	\\
2008 Nov 02 & Scl4 S	& 01:00:03.96 	& $-$33:15:30.60	& B	& 300	& 1.7	& 1.05-1.08	\\
2008 Nov 02 & 		&		&		& V	& 300	& 1.9 	& 1.06-1.09	\\
2008 Nov 02 & 		&		&		& I	& 600	& 1.5	& 1.07-1.10	\\
2008 Nov 02 & Scl5 S	& 01:00:03.96 	& $-$34:25:30.36	& B	& 300	& 1.5	& 1.17-1.23	\\
2008 Nov 02 & 		&		&		& V	& 300	& 1.5 	& 1.18-1.27	\\
2008 Nov 02 & 		&		&		& I	& 600	& 1.4	& 1.21-1.29	\\
2008 Nov 02 & Scl6 S	& 01:05:40.20 	& $-$33:50:03.12	& B	& 300	& 2.0	& 1.92-2.18	\\
2008 Nov 02 & 		&		&		& V	& 300	& 1.8 	& 1.97-2.30	\\
2008 Nov 02 & 		&		&		& I	& 600	& 1.9	& 2.07-2.39	\\
2008 Nov 01 & Scl8 N	& 01:00:03.96 	& $-$33:22:31.08	& B	& 300 	& 1.5	& 1.05-1.08	\\
2008 Nov 01 & 		&		&		& V	& 300	& 1.4 	& 1.04-1.07	\\
2008 Nov 01 & 		&		&		& I	& 600	& 1.4	& 1.03-1.06	\\
2008 Nov 01 & Scl9 N	& 00:57:16.92 	& $-$32:57:23.40	& B	& 300 	& 1.8	& 1.00-1.01	\\
2008 Nov 01 & 		&		&		& V	& 300	& 1.7 	& 1.00-1.01	\\
2008 Nov 01 & 		&		&		& I	& 600	& 1.4	& 1.00-1.01	\\
2008 Nov 01 & Scl10 N	& 00:57:14.40 	& $-$34:07:22.80	& B	& 300 	& 1.6	& 1.01-1.03	\\
2008 Nov 01 & 		&		&		& V	& 300	& 1.5 	& 1.02-1.04	\\
2008 Nov 01 & 		&		&		& I	& 600	& 1.4	& 1.02-1.04	\\
2008 Nov 01 & Scl11 N	& 01:02:53.16 	& $-$34:07:23.52	& B	& 300 	& 1.5	& 1.05-1.11	\\
2008 Nov 01 & 		&		&		& V	& 300	& 1.5 	& 1.07-1.12	\\
2008 Nov 01 & 		&		&		& I	& 600	& 1.3	& 1.10-1.14	\\
2008 Nov 01 & Scl12 N	& 01:02:51.00 	& $-$32:57:24.12	& B	& 300 	& 1.3	& 1.26-1.33	\\
2008 Nov 01 & 		&		&		& V	& 300	& 1.4 	& 1.27-1.35	\\
2008 Nov 01 & 		&		&		& I	& 600	& 1.4	& 1.30-1.38	\\
\hline 
\end{tabular}
\end{center}
\end{table*}
The 8 chips of each MOSAIC image suffer from electronic crosstalk which causes ghost images. Each pair of CCD's are controlled by one CTIO Arcon controller, which causes one CCD to show ghost images from stars in a different CCD that are saturated or nearly saturated. This crosstalk was removed using correction terms provided by NOAO. \\
The flat-field obtained for the MOSAIC camera is made up of different chips. Instead of normalizing each chip individually, all chips were normalized using the average mean across the whole image. The normalization factor for each chip then depends on its exact gain, and can be used to correct the science images both for the variation of sensitivity across each chip and to normalise the gain between different chips to a common value, thus obtaining a final image which has one uniform average gain. The standard flat-field obtained using the dome lamp was improved with the use of sky flat-fields. Sky flat-fields better approximate uniform illumination than possible with the dome lamp, as light enters the telescope in the same way as during science observations. The sky flat-field provides a correction on the dome flat-field, creating a master flat-field which is used to correct the science images. The MOSAIC I-band images also suffer from fringing, caused by constructive and destructive interference of night sky lines reflected in the CCD substrate. These fringe patterns were removed using an archival I-band fringe frame. \\
We created a bad pixel mask to flag saturated and negative pixels, cosmic rays and any damaged areas on each CCD. Cosmic rays were detected with a moving 7$\times$7 pixel block average filter with the central target pixel excluded. The average value of the block is a prediction for the value of the target pixel, and the median of a background annulus around the block is used to protect cores of stars from being flagged as bad pixels. If the difference between the average block value and the central pixel value was larger than a threshold, the target pixel was flagged as a cosmic ray and added to the bad pixel mask. Subsequently, we remove these bad pixels from the image by interpolating over them, preventing them from propagating into adjacent pixels during the resampling process. \\
Due to the radial distortion in the large field of view the pixel scale is not constant across the field-of-view. Therefore the images had to be resampled to a common uniform pixel scale before stacking different exposures that have been offset with respect to each other because of telescope pointing errors between multiple observations of the same field. This requires a precise World Coordinate System (WCS), which provides the rotation, scale and offsets for mapping pixel coordinates to sky coordinates. The WCS header assigned to each image by the CTIO 4m-Blanco telescope was not sufficient to accurately align the images. Therefore, we obtained a new WCS solution using stars from the 2MASS point source catalog~\citep{Skrutskie06} that were found in our field-of-view. The solution uses a ZPN projection~\citep{Calabretta02}, which was fitted on each chip individually, while taking into account it's position with respect to a reference point (the center of the mosaic). \\
From a comparison of the final calibrated positions with 2MASS positions we find that the WCS is accurate to $\pm$0.2$^{\prime\prime}$ across the entire field of view. This level of accuracy~(less than one pixel) is sufficient to align individual images. The images are resampled to a uniform pixel scale of 0.27$^{\prime\prime}$ and then scaled to a common exposure time and sky level and stacked together to produce as deep a final image of each pointing as is possible. 
\begin{table*}[!ht]
\caption[]{Photometric solution parameters and uncertainties. \label{photpars}}
\footnotesize
\begin{center}
\begin{tabular}{cccc}
\hline\hline
Band \& colour	& $A_{mag}$	&  $\alpha$		& $\beta$	\\
\hline
\multicolumn{4}{c}{4m} \\
\hline
B,B-V & 5.133$\pm$0.009	& 0.246$\pm$0.001 & $-$0.077$\pm$0.001		\\
I,V-I &	5.501$\pm$0.001	& 0.062$\pm$0.001 & 0.031$\pm$0.001	 	\\
V,B-V &	4.917$\pm$0.001	& 0.127$\pm$0.001 & 0.049$\pm$0.001		\\
V,V-I & 4.924$\pm$0.002	& 0.129$\pm$0.001 & 0.030$\pm$0.001		\\
\hline 
\multicolumn{4}{c}{0.9m} \\
\hline
B,B-V & 7.969$\pm$0.013	& 0.309$\pm$0.006 & 0.069$\pm$0.003		\\
I,V-I &	8.749$\pm$0.011	& 0.031$\pm$0.005 & $-$0.028$\pm$0.002	 	\\
V,B-V &	7.936$\pm$0.013	& 0.084$\pm$0.007 & $-$0.039$\pm$0.007		\\
V,V-I & 7.928$\pm$0.012	& 0.084$\pm$0.007 & $-$0.046$\pm$0.006		\\
\hline 
\end{tabular}
\end{center}
\end{table*}

\subsection{CTIO 0.9m}
The 0.9m telescope is equiped with a single 2048$\times$2046 pixel CCD detector, which has a field of view of 13.5$^{\prime}$ $\times$ 13.5$^{\prime}$ and a pixel scale of 0.396$^{\prime\prime}$. We use standard bias and flat-field removal techniques in IRAF for the reduction. A predefined bad pixel mask provided by NOAO is used to flag bad pixels, since the CCD's of the 0.9m telescope have stable defects. A precise WCS is obtained, in order to match all stars observed in the 0.9m images to the same stars observed in the 4m images. As for the 4m data the WCS solution is derived using stars from the 2MASS catalog, using a ZPN projection of sky coordinates to pixel coordinates. The final solution is also accurate to $\pm$0.2$^{\prime\prime}$ across the image, which corresponds to roughly half a pixel. This is sufficiently accurate to cross-correlate the 0.9m source catalogues with those of the 4m.

\section{Photometry}
\label{photometry}
Photometry is performed on all the reduced images~(0.9m and 4m) using DoPHOT~\citep{Schechter93}, using a version which is optimized to carry out accurate photometry down to the faintest levels. The methodology used in this section is described in detail in~\citet{Saha10}. \\
DoPHOT uses an analytical function as a model PSF for describing different object types. Each model is characterized by six different parameters~(X position, Y position, total intensity and three shape parameters), used both for the detection of objects and for determining their brightness. The initial shape parameters for star-like objects are determined separately for each image using a sample of bright, isolated stars. These parameters are updated after each consecutive pass through the image. 
\begin{figure*}[!htb]
\centering
\includegraphics[angle=0, width=0.9\textwidth]{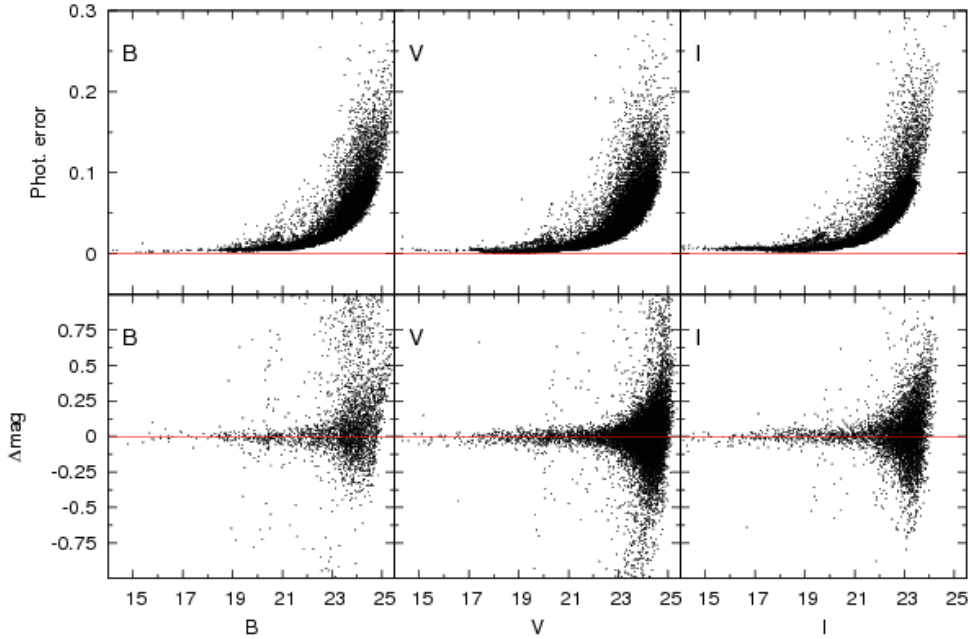}
\caption{\textbf{Top row:} Calibrated PSF magnitudes plotted against photometric errors for the central pointing for B, V \& I filters. \textbf{Bottom row:} Magnitude difference between stars in overlapping regions of different pointings, for B,V \& I filters.  \label{Sclerror}} 
\end{figure*}
\\
The model PSF is not allowed to vary with position on the image. Instead, a fixed PSF is used, which is only allowed to vary from chip to chip. However, as the true PSF is expected to vary across the image~(and also across each chip), we incorporate a position-dependency into the determination of the aperture corrections. Aperture magnitudes~(which are independent of the PSF shape) are determined for bright isolated stars across each chip. The difference between the PSF and aperture magnitudes is mapped as a function of position over the field of view, and this provides the correction factors for DoPHOT PSF magnitudes. \\
The fitting function used to map the difference between aperture and PSF magnitudes is different for 4m and 0.9m images. In the case of the 0.9m images, a simple quadratic surface is used, to allow variations across the image. In the case of the MOSAIC images a more complex surface is used, to allow variations across each chip to be treated independently, in addition to a quadratic surface across the whole image to fit tilts or offsets between the individual chips~(See~\citet{Saha10} for details). \\
The photometry of the short and long exposures is combined by first cross-correlating the stellar positions. Subsequently, for stars that have are found in both exposures the star with the smallest photometric error is used in the combined catalog. Stars with bright magnitudes that were found only in the short exposure are also added to the catalog. 

\subsection{Photometric Calibration}
\label{calibration}
We use standard star observations of fields taken from~\citet{Landolt92, Landolt07} with included Stetson standard stars~\citep{Stetson00} to obtain photometric solutions for each night. The photometry is calibrated to the Johnson-Cousins-Glass photometric system using a photometric solution of the form:
\begin{displaymath}
 mag_{obs} = mag_{true} + A_{mag} + \alpha * \chi + \beta * colour
\end{displaymath}
where $mag_{obs}$ is the observed magnitude, $mag_{true}$ the true magnitude, $\chi$ the observed airmass, colour the observed colour index used, $A_{mag}$ the zeropoint correction term, $\alpha$ the airmass coefficient and $\beta$ the colour term. \\
We determined the photometric solutions automatically with a Bayesian approach, which provides a statistically robust method to separate reliable measurements from inaccurate ones. We used a Monte Carlo method to determine the most probable photometric solution, given the standard star measurements and taking into account the possibility of an unknown fraction of inaccurate measurements. \\
Using this method, we obtain likelihood distributions for the coefficients of the photometric calibration, the fraction of outliers and the parameters of the outlier distribution. Furthermore, the likelihood of being an outlier is given for each standard star measurement. The mean of the likelihood distribution is adopted as the photometric solution, with the uncertainty of each coefficient given by the standard deviation of the corresponding distribution. The resulting photometric coefficients, along with their uncertainties are given in Table.~\ref{photpars}. A detailed description of the checks done to ensure a reliable photometric calibration is given in Appendix~\ref{calappendix}. \\
The resulting catalogs of different bands are matched together to create one full catalog containing 147456 stars in the V band down to 25.5 mag, 72577 in the B band going down to 25.7 mag and 116414 in the I band going down to 24.3. The top panels of Fig.~\ref{Sclerror} show the DoPHOT internal errors of all these stars as a function of their magnitude for the central pointing only. The other pointings display similar profiles, with the exception of the outer B band observations, which have shorter integration times and thus go to different depths. At the bright end the best error is $\approx$0.002, while at the faint end the error is $\approx$0.2 for the faintest MSTO stars at the magnitude levels quoted above. The catalog contains only the stars for which all PSF parameters could be determined~(type 1) and those that are found as mergers of two stars~(type 3).  \\
The photometry of the outer pointings is placed on the same system as the central pointing by using stars in the overlapping region. The multiple measurements are used to fit an offset and colour term which put the photometry on the same system as the central pointing. Comparison of the magnitudes of stars which have two or more measurements in the overlapping regions are shown in the bottom panels of Fig.~\ref{Sclerror}. The figure shows that the photometry of all pointings is calibrated to the same scale as the central pointing within the uncertainty of the photometric measurements. The photometric accuracy of the final catalog coming from random errors is $\approx$0.002 mag for the brightest stars, going down to $\approx$0.2 mag for the faint stars.

\subsection{Artificial star tests}
\begin{figure}[!h]
\centering
\includegraphics[angle=270, width=0.485\textwidth]{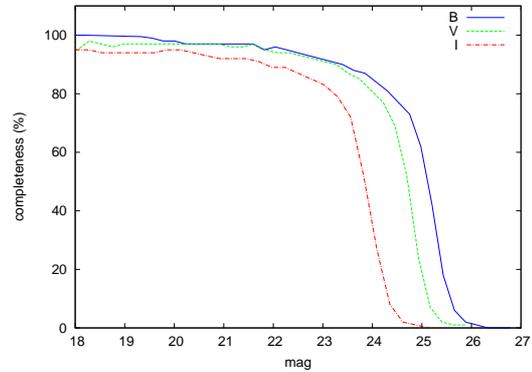}
\caption{Completenes fractions of the central Sculptor field as a function of magnitude for the B,V and I filter. For stars brighter then shown here the completeness factor is assumed to be 100\%. \label{compplot}} 
\end{figure}

Given that our photometry only just extends beyond the magnitude of the oldest MSTOs it is important to quantify the completeness down to the faintest magnitudes. Incompleteness is caused by photometric errors, crowding effects and the detection limit of the images, and results in the loss of stars at faint magnitudes. In order to derive the completeness fraction at different magnitudes we have performed artificial star tests on the deep stacked images. Tests are done by inserting artificial stars in the images (a number of the order of 5\% of the total) and re-doing the photometry in the same way. The resulting catalog is matched to the observed and artificial input catalogs to obtain the fraction of stars recovered~(and their measurement errors) at different magnitudes and colours. The completeness fraction for different filters is plotted for the central Sculptor pointing in Fig.~\ref{compplot} as a function of magnitude. For the other pointings the completeness profile looks similar in the V and I filters. Due to bad seeing conditions the B completeness profile in the outer pointings drops off faster than shown in Fig.~\ref{compplot}. For magnitudes brighter than shown in the figure the catalogs are assumed to be complete. Fig.~\ref{compplot} shows that the 50\% completeness level is located at B$\approx$25.2, V$\approx$24.8 and I$\approx$24.0, which means our limiting magnitudes are about 1 magnitude deeper than the oldest MSTOs.  

\subsection{The Colour-Magnitude Diagrams}
\label{description}
\begin{figure*}[!htb]
\centering
\includegraphics[angle=0, width=0.9\textwidth]{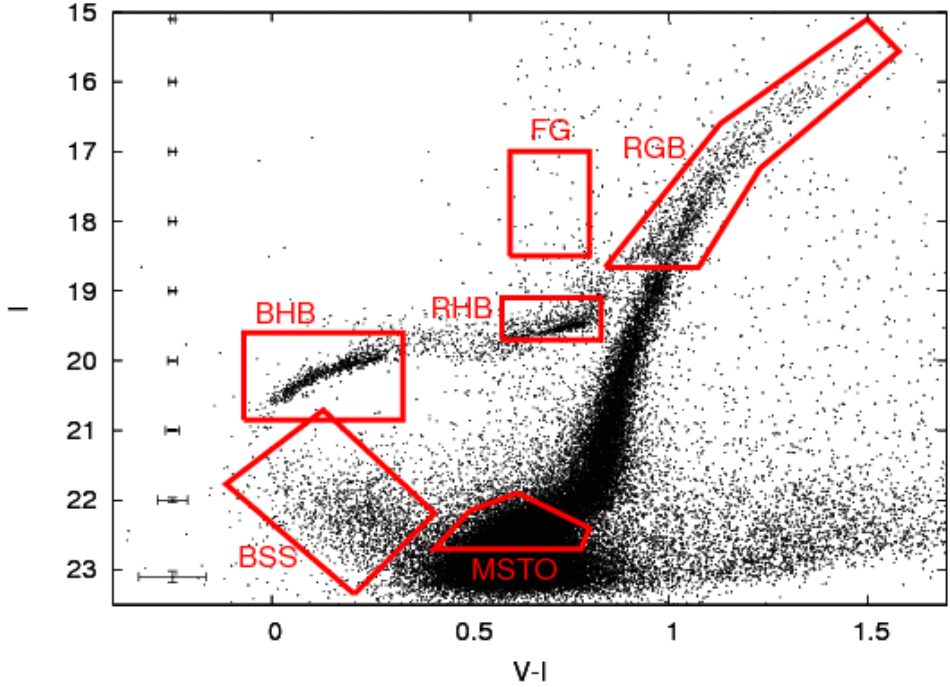}
\caption{Sculptor (I,V$-$I) colour magnitude diagram of the central region overlaid with regions identifying the different stellar populations. BHB: Blue Horizontal Branch, RHB: Red Horizontal Branch, RGB: Red Giant Branch, BSS: Blue Straggler Stars, MSTO: Main Sequence Turn-Off. In the RGB box numerous AGB stars can also be observed. A box is also included which provides an estimate of the contamination by foreground stars from the Milky Way~(FG). Photometric errors are shown as the size of the errorbar at the corresponding magnitude level. \label{SclVIboxes}} 
\end{figure*}

\begin{figure*}[!ht]
\centering
\includegraphics[angle=0, width=0.9\textwidth]{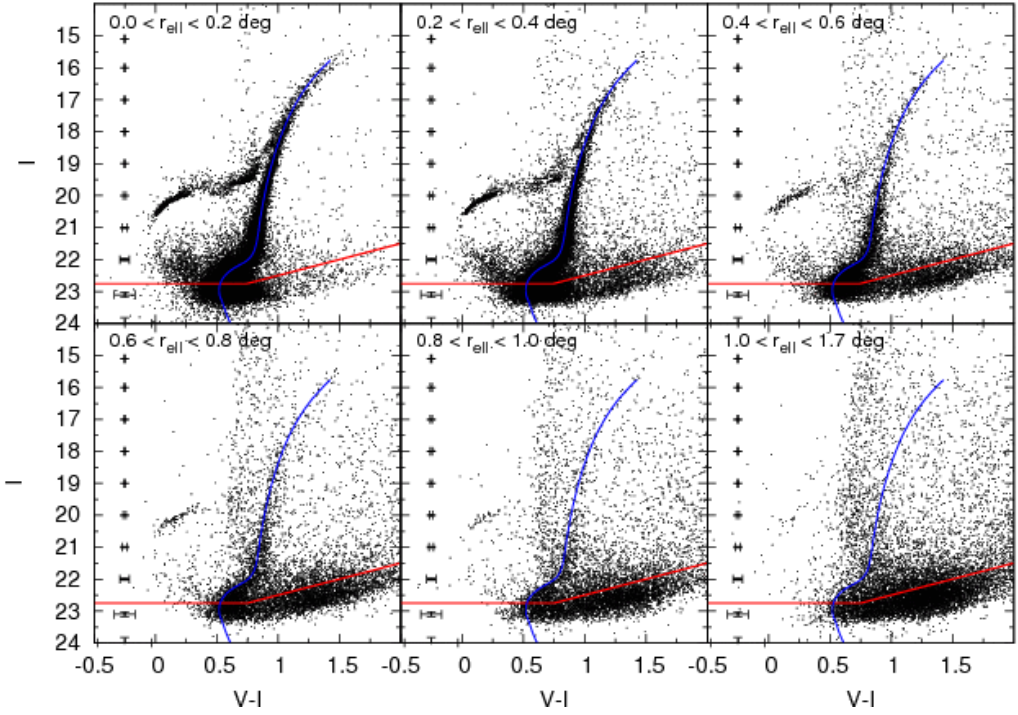}
\caption{(I,V$-$I) Colour-Magnitude Diagrams of the Sculptor dSph for increasing elliptical radius. Error bars show the average photometric error at that level. The 50$\%$ completeness level for this subsample is indicated by the (red) solid line. the blue line shows a reference isochrone~([Fe/H]=$-$1.80,[$\alpha$/Fe]=0.2, age = 11 Gyr) which runs through the middle of the RGB in the centre. \label{SclVIrad}} 
\end{figure*}

\begin{figure*}[!ht]
\centering
\includegraphics[angle=0, width=0.9\textwidth]{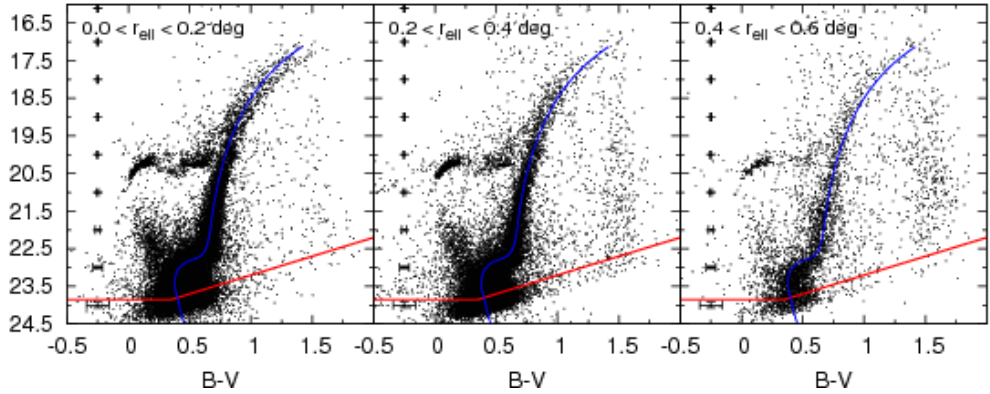}
\caption{(V,B$-$V) Colour-Magnitude Diagrams of the Sculptor dSph for increasing elliptical radius. Error bars showing the average photometric error and a 50$\%$ completeness line~(in red) are also included. A reference isochrone indicating the middle of the RGB in the centre is also shown as a blue line. \label{SclBVrad}} 
\end{figure*}

\begin{figure}[!ht]
\centering
\includegraphics[angle=270, width=0.485\textwidth]{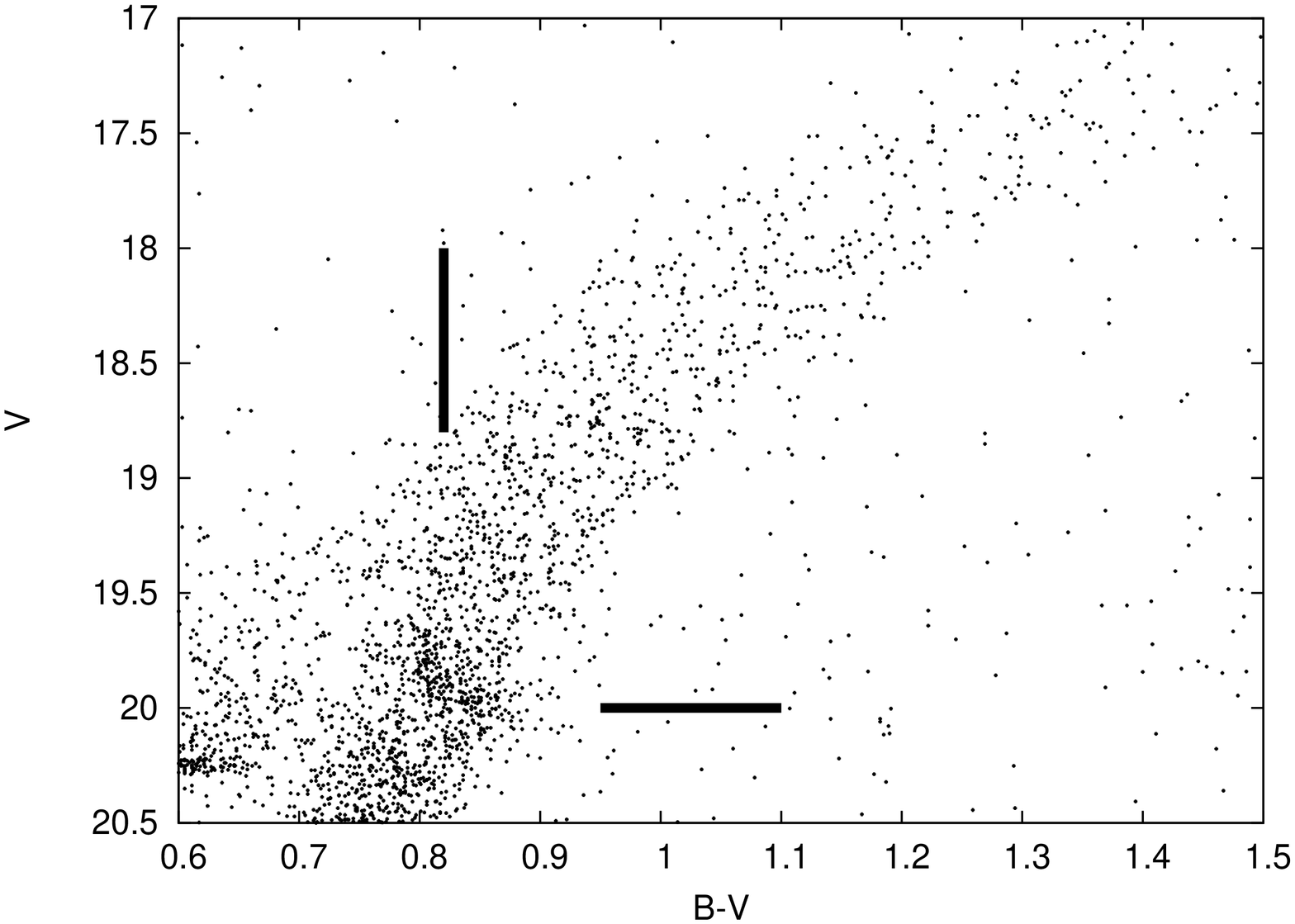}
\caption{Zoom in of the RGB bump region in the (V,B$-$V) CMD of the Sculptor dSph. A composite RGB bump is visible at 0.78$<$B-V$<$0.88 and 19.6$<$V$<$20.1.  \label{SclRGBzoom}} 
\end{figure}

Using the full calibrated photometry catalog we can now produce Colour-Magnitude Diagrams~(CMDs) to study different evolutionary features of the Sculptor dSph. Fig.~\ref{SclVIboxes} shows an (I,V$-$I) CMD of the inner part of Sculptor, where we have identified the different evolutionary features that can be seen in all the CMDs.
\\
Figs.~\ref{SclVIrad} and~\ref{SclBVrad} show calibrated (I,V$-$I) and (V,B$-$V) CMDs for the Sculptor dSph for different annuli of elliptical radius~(r$_{ell}$). A line indicating the 50\% completeness levels of these stars is also included. The elliptical radius is defined as the major axis of the ellipse centred on Sculptor~(RA=01:00:09, DEC=$-$33:42:30 with ellipticity e=0.32, see~\citet{Mateo98}). We assume a tidal radius of 76.5 arcmin and core radius of 5.8 arcmin~\citep{Irwin95}. No de-reddening is applied to the CMDs in Figs.~\ref{SclVIrad} and~\ref{SclBVrad}; Instead, models and isochrones used to analyse the CMDs are reddened using the same extinction coefficient as obtained for the Sculptor dSph. \\
The CMDs show only the stars for which all PSF shape parameters are determined. Therefore, the CMDs do not go as deep as the limits determined from the completeness profile~(Fig.~\ref{compplot}). If we consider the stars for which not all the shape parameters have been determined as 'lost', the 50\% completeness levels for this subsample of stars are B$\approx$23.9, V$\approx$23.5 and I$\approx$22.8. We stress that this does not mean that 50\% of the stars at these levels are not observed, but only that these are not considered in the subsequent analysis. For the (I,V$-$I) CMDs full coverage is obtained for radii out to r$_{ell}$=0.8 deg and for the (V,B$-$V) CMDs out to r$_{ell}$=0.6 deg. 
\\
The CMDs display several features which remain clearly visible out to r$_{ell}$=1 deg~(such as the RGB and BHB). The RHB, however, disappears around r$_{ell}$=0.6 deg where the number of foreground stars becomes comparable to the number of RHB stars. A population of potential Blue Straggler Stars~(BSS) remains visible  out to r$_{ell}$=1 deg. The old MSTO stars remain visible at all radii, but are strongly affected by Milky Way foreground contamination beyond r$_{ell}$=1 deg. The RGB has a spread in colour which is indicative of a spread in metallicity and/or age. A composite RGB bump is also present, for which a zoom in of the CMD is presented in Fig.~\ref{SclRGBzoom}. This is consistent with the presence of an extended period of star formation.
\\
To estimate the effect of foreground stars in the CMDs, we have used the Besan\c{c}on models~\citep{Robin03} to construct a model Galactic population in the direction of the Sculptor dSph. A comparison of the number of foreground stars in the CMD~(FG, see Fig.~\ref{SclVIboxes}) with the predicted foreground stars of the same colours from the Besan\c{c}on models agree to within 10$\%$, validating the use of the Besan\c{c}on models to accurately predict the number of foreground stars anywhere in the CMD. \\
Besides foreground stars a cloud of faint red stars is seen close to the detection limit in the (I,V$-$I) CMD, redward of V$-I\approx$1.0, which increases in number as we move out of the centre of Sculptor. These objects are consistent with the integrated colours of background spiral and irregular galaxies at a redshift z=0.5-0.8~\citep{Fukugita95} that remain unresolved due to a lack of spatial resolution and sensitivity. \\
In order to disentangle the positions of the different contaminating populations present in our CMDs, we have constructed a Colour-Colour diagram~({Fig.~\ref{Sclcolcol}a). The prominent evolutionary phases and foreground and background contamination have been colour-coded. The position of the same populations are shown in the (I, V$-$I) CMD (Fig.~\ref{Sclcolcol}b), and the (V, B$-$V) CMD (Fig.~\ref{Sclcolcol}c).
\begin{figure}[!ht]
\centering
\includegraphics[angle=0, width=0.45\textwidth]{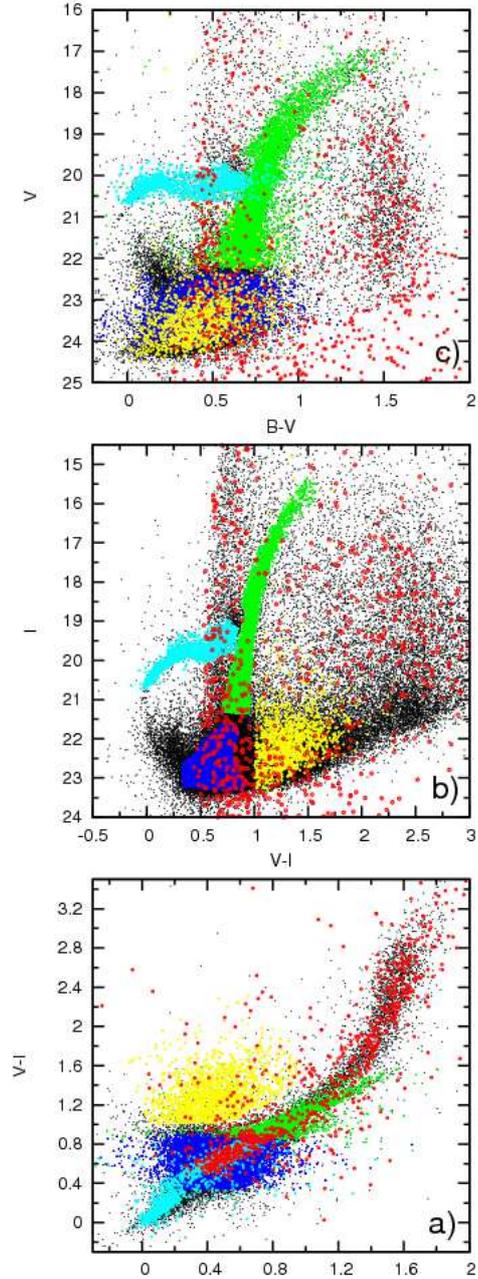}
\caption{\textbf{a)} Sculptor (B$-$V),(V$-$I) colour-colour diagram~(black points) overlaid with colour-coded CMD regions: Horizontal Branch~(light-blue), RGB/SGB~(green), MSTO~(dark blue), MW foreground~(red) and unresolved galaxy background~(yellow). \textbf{b)} (V,B$-$V) and (I,V$-$I) (\textbf{c}) colour magnitude diagrams with the same colour-coding. \label{Sclcolcol}} 
\end{figure}
\\
It is clear from Fig.~\ref{Sclcolcol} that the Milky Way foreground contaminates several features in the Sculptor CMDs, including the MSTO. The foreground stars become distinct from the main Sculptor population in Colour-Colour space for the reddest colours, as can be seen from the distinct 'plume' of stars in Fig.~\ref{Sclcolcol}a at (V$-$I)$\approx$1.6-3.4 and (B$-$V)$\approx$1.2-1.8. The unresolved background galaxies are most distinct from the Sculptor MSTO population in (V$-$I), making them show up as a separate 'cloud' of red stars, while in (B$-$V) they overlap with the MSTO. Therefore, having three filters helps in the interpretation of the CMDs. CMDs using different combinations of filters are also differently sensitive to age and metallicity effects~(de Boer et al., 2011, in prep). 

\subsection{Structural Parameters}
\label{structure}
Having quantified the most prominent sources of foreground and background contamination in our CMDs we can now investigate the spatial distributions of the different CMD features using star counts in foreground/background corrected regions. We map the spatial distributions of different populations in the Sculptor dSph using 2D iso-density contour maps. The stars are convolved with a normalized Gaussian with a width of $\sigma$=0.05 deg to generate intensity distributions. These distributions are then used to construct iso-density contours. We obtain the intensity level of the foreground stars by applying the same procedure to the Besan\c{c}on model predictions, spread homogeneously across the field. 
\begin{figure*}[!ht]
\centering
\includegraphics[angle=0, width=0.45\textwidth]{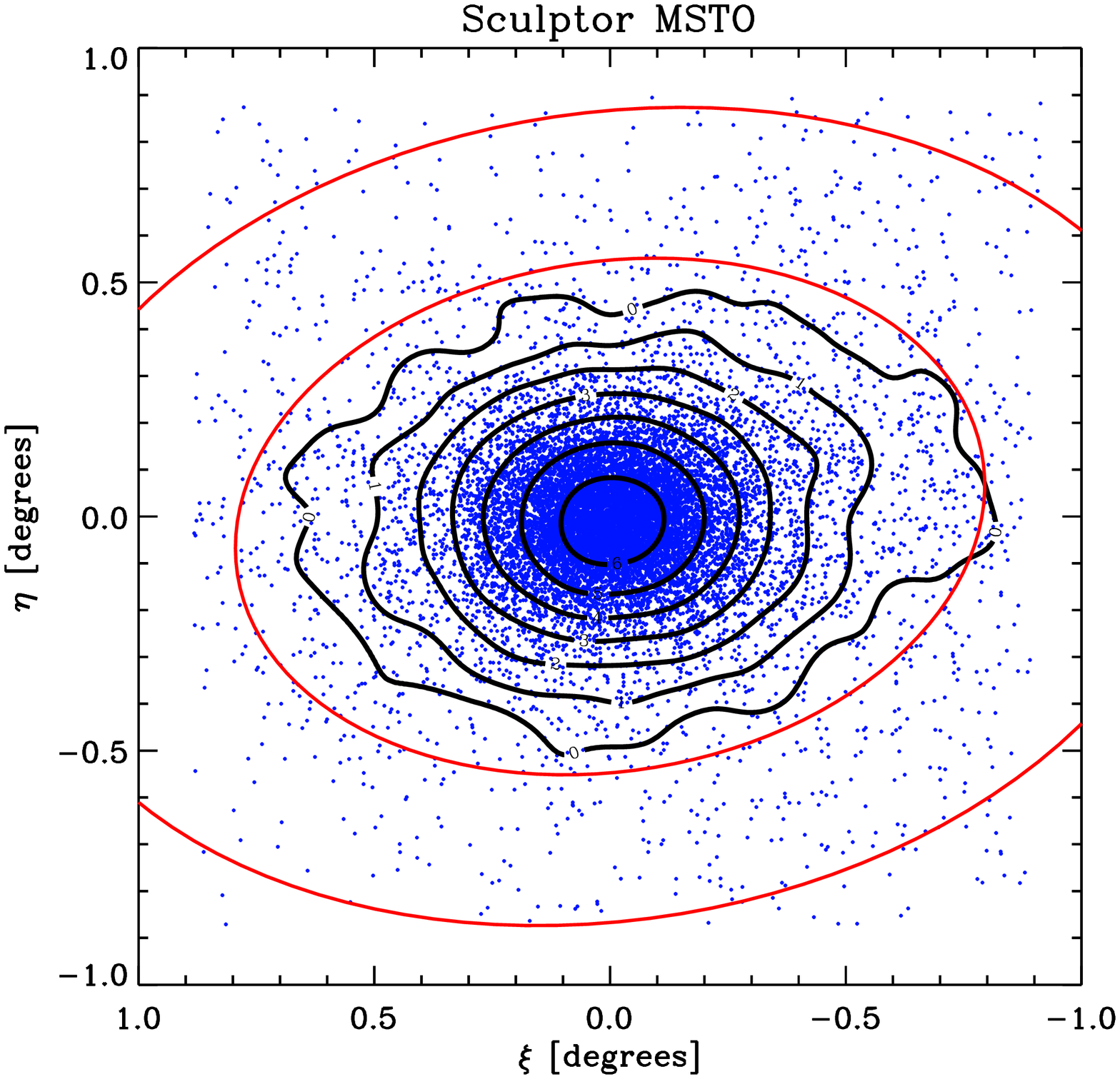}
\includegraphics[angle=0, width=0.45\textwidth]{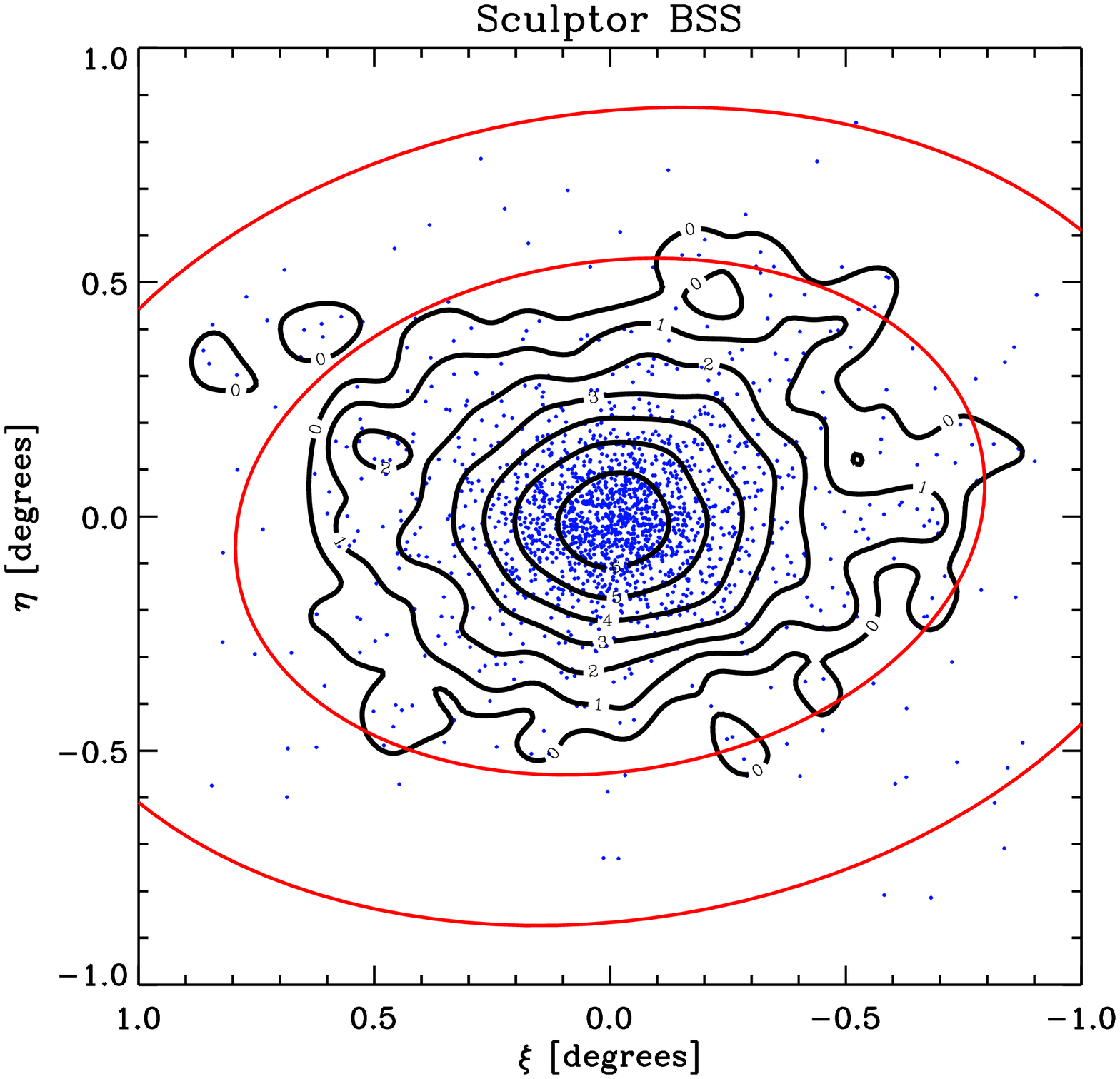}
\caption{Contour maps of the MSTO and BSS populations overlaid on the individual stars, shown as dots. The outer (red) ellipse is the tidal radius of the galaxy. The inner (red) ellipse is the largest ellipse not influenced by contour deformation caused by the proximity of the edge of observational coverage. The outermost contour in each plot is defined as 1$\%$ of the foreground corrected maximum intensity level. Subsequent contour levels increase by a factor 2 for each contour. North is up, and East is to the left. \label{contours}} 
\end{figure*}
\\
Since the centre of the resulting contour maps was found to be off-centre with respect to the published RA and DEC of the Sculptor centre we decided to re-derive the structural parameters of Sculptor using our current data set. Using 2D spatial Hess diagrams with a bin size of 0.02$\times$0.02 deg we investigated the radial variation of the central position, eccentricity and position angle with the IRAF task \textbf{ELLIPSE}. The analysis reveals that the central position of Sculptor in our data set is located at RA=01:00:06.36, DEC=$-$33:42:12.6, which is west~($-$0.66$^{\prime}$$\pm$0.06$^{\prime}$) and north~(0.288$^{\prime}$$\pm$0.05$^{\prime}$) of the central coordinates listed by~\citet{Mateo98}. The eccentricity is found to be e=0.26$\pm$0.01 and the position angle PA=85.26$\pm$0.91, compared to the literature values of e=0.32 and PA=99~\citep{Irwin95}. These parameters for our data set are adopted for the rest of the analysis in this paper. \\
With these parameters the contour maps of the bright stars~(down to the HB) confirm previous results, showing that the outer contours are more elliptical than the inner and that the spatial distribution of the RHB and BHB populations are different. We show for the first time the spatial distribution maps of the MSTO and BSS population in Fig.~\ref{contours}. The BSS population looks similar in extent to the RGB and BHB. It shows the same change in ellipticity of the contours with different radii, as observed in the brighter features. \\
A small overdensity is visible around $\xi$=$-$0.5, $\eta$=0.1 in the MSTO stellar distribution in Fig.~\ref{contours}, which is more prominent in the spatial Hess diagram~(Fig.~\ref{MSTOhess}). The photometric overdensity has about 30$\pm$5.5 stars per bin of 4.5$\times$4.5 arcmin, compared to the mean number of stars of 13.6$\pm$3.7 per bin at the corresponding elliptical radius. The feature coincides spatially with the stars of a velocity substructure found in ESO/FLAMES spectroscopy~\citep{Battaglia07}, which are overlaid on the Hess diagram in Fig.~\ref{MSTOhess}. The velocity substructure is made up of stars found in a cold structure at 0.2 $<$ r$_{ell}$ $<$ 0.6 deg, with heliocentric velocities between 128 and 142 km/s and a velocity dispersion of 2.4$\pm$0.7 km/s. The spectroscopic metallicities cover a range of $-$2.8$<$[Fe/H]$<-$1.6. The substructure can be split up into two parts, one at 0.2 $<$ r$_{ell}$ $<$ 0.3 deg~(crosses in Fig.~\ref{MSTOhess}) with a narrow metallicity range of [Fe/H]=$-$1.96$\pm$0.05 and one at 0.3 $<$ r$_{ell}$ $<$ 0.6 deg~(circles in Fig.~\ref{MSTOhess}) with a broad metallicity range of $-$2.8$<$[Fe/H]$<-$1.6. The photometric overdensity matches the spatial position of the velocity substructure at 0.3 $<$ r$_{ell}$ $<$ 0.6 deg. In terms of the CMD, the stars in the substructure match the same distribution as the main body of Sculptor stars, indicating that they are likely drawn from the same population, with a broad metallicity range. \\
In order to investigate asymmetries in the spatial distributions of the different CMD populations we performed starcounts within annuli centred on Sculptor. For different r$_{ell}$ the fraction of stars West~($\xi$$<$0) of the centre versus East~($\xi>$0) of the centre is measured within an annulus. Fig.~\ref{Sclasym} shows the fraction N$_{West}$/N$_{East}$ for the full CMD~(green circles), a selection of the CMD containing only Sculptor~(black crosses), BSS population~(blue asterisks), foreground stars~(cyan filled boxes) and the unresolved background galaxies~(red plusses). 
\begin{figure}[!htb]
\centering
\includegraphics[angle=270, width=0.5\textwidth]{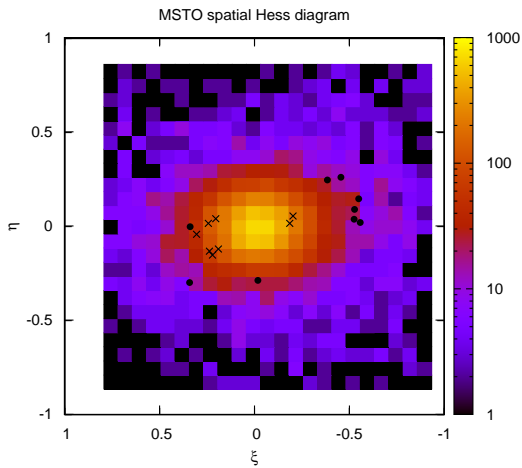}
\caption{Spatial Hess diagram of the MSTO population centred on Sculptor. Overlaid on the diagram are stars found in a velocity substructure~(crosses indicate the substructure at 0.2 $<$ r$_{ell}$ $<$ 0.3 deg and circles the one at 0.3 $<$ r$_{ell}$ $<$ 0.6 deg.). A small photometric overdensity is visible at $\xi$=-0.5, $\eta$=0.1, which coincides with one of the velocity substructures. \label{MSTOhess}} 
\end{figure}
\\
The Sculptor populations~(overall Sculptor, BSS) are almost symmetric for small elliptical radii, becoming more asymmetric further out. The foreground stars show a symmetric distribution at all radii, consistent with Galactic population models. The unresolved background galaxies~(see Fig.~\ref{Sclcolcol}) display an anti-symmetric distribution with regards to the Sculptor features, which might be caused by obscuration of background galaxies by Sculptor stars, or a position-dependent morphological misclassification of galaxies and stars. The asymmetry difference of BSS stars with regards to the overall Sculptor population discussed by~\citet{Mapelli09} is not reproduced here and is likely due to the comparison with the full CMD at large radii instead of with the overall Sculptor population. The figure shows that the overall Sculptor population displays a slight asymmetry, with $\approx$10\% more stars in the West than in the East, in the same direction as observed earlier by~\citet{Eskridge882}.
\begin{figure}[!ht]
\centering
\includegraphics[angle=270, width=0.485\textwidth]{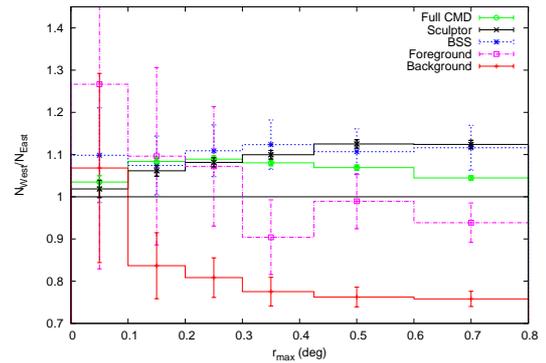}
\caption{Fraction of Sculptor stars in the West~($\xi$$<$0) vs East~($\xi$$>$0) for different annuli of elliptical radius. Different populations are indicated with different colors,  full CMD~(green circles), overall Sculptor~(black crosses), BSS~(blue asterisks), foreground~(cyan filled boxes) and background galaxies~(red plusses) \label{Sclasym}} 
\end{figure}

\section{Interpretation}
\label{interpretation}
Using the calibrated photometric catalog we can look at the behaviour of any stellar population over a large fraction of the galaxy. We can study the radial distributions of different CMD features and search for variations with elliptical radii. By analysing the CMDs of the MSTO region and combining this with information coming from spectroscopic surveys we can quantify the ages and metallicities of the different populations making up the Sculptor dSph. 

\subsection{The Horizontal Branch}
\label{int:HB}
We know from previous studies of the Sculptor dSph~\citep{Majewski99, HurleyKeller99,Tolstoy04} that two distinct HB populations are present, with different radial distributions. This can also be seen in our data in Fig.~\ref{radhistHB}, showing the foreground corrected number of stars in each bin of elliptical radius in the RHB and BHB populations. Since the number of foreground stars is not predicted to vary significantly with position across Sculptor the foreground correction is made by taking the predicted number of Galactic stars per unit area~(as given by the Besan\c{c}on models) within the same magnitude and colour range as the CMD feature. This number is then multiplying by the area of the ellipses of different r$_{ell}$, and subtracted from the observed star counts. The figure shows that the RHB population dominates in the central region, after which it falls off more rapidly than the BHB distribution. The BHB population is more uniformly distributed across the galaxy, and dominates only in the outer region.
\begin{figure}[!ht]
\centering
\includegraphics[angle=270, width=0.45\textwidth]{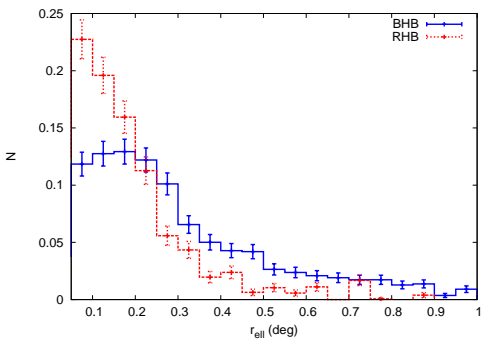}
\caption{Radial histograms of RHB~(red, dashed) and BHB~(blue, solid) populations in the Sculptor dSph. \label{radhistHB}} 
\end{figure}
\\
Besides the spatial distribution, the HB of Sculptor has also been extensively studied to look for variable stars. More than 90 RR Lyrae stars have been found in Sculptor, with average [Fe/H]$\approx-$1.7 or lower along with 3 anomalous metal-poor Cepheids~\citep{Kaluzny95}. This confirms that the bulk of the stellar population in Sculptor is metal-poor. In addition, 2 more metal-rich short period variables (with [Fe/H]$\le-$0.7) were found within the core radius, in accordance with the spatial distribution of the two HB populations.

\subsection{The Red Giant Branch}
\label{int:RGB}
The RGB of Sculptor is long known to have a larger colour spread than seen in Globular Clusters~\citep{DaCosta84}, consistent with it having experienced an extended period of star formation. However, no radial differences as clear as those on the HB have been found. This is because all populations overlap on the RGB, especially at low metallicity, and are thus not clearly distinguishable from each other. \\
In Fig.~\ref{radRGB} we show the RGB region in the (I,V$-$I) CMD for different elliptical radii. Overlaid on the CMDs are two isochrones from the Dartmouth Stellar Evolution Database~\citep{DartmouthI}. The isochrones are representative of the most metal-poor~([Fe/H]=$-$2.45) and most metal-rich~([Fe/H]=$-$0.90) stars making up the bulk of the population of Sculptor, as derived from RGB spectroscopy~\citep{Tolstoy04, Battaglia07}. The most metal poor stars found in Sculptor with these surveys have a metallicity of [Fe/H]$\le-$2.5, which is not available in the Dartmouth Stellar Evolution Database. Instead, the most metal poor isochrone available in the library~([Fe/H]=$-$2.45) is used. The ages of the isochrones have been selected to fit the colours of RGB stars of corresponding metallicity. For increasing elliptical radius it can be seen that the brightest and reddest part of the RGB diminishes, as does the blue side of the Subgiant Branch. From the overlaid isochrones it can be seen that this is broadly consistent with the disappearance of the metal-rich component, as expected from the HB distributions. 
\begin{figure*}[!htb]
\centering
\includegraphics[angle=270, width=1.0\textwidth]{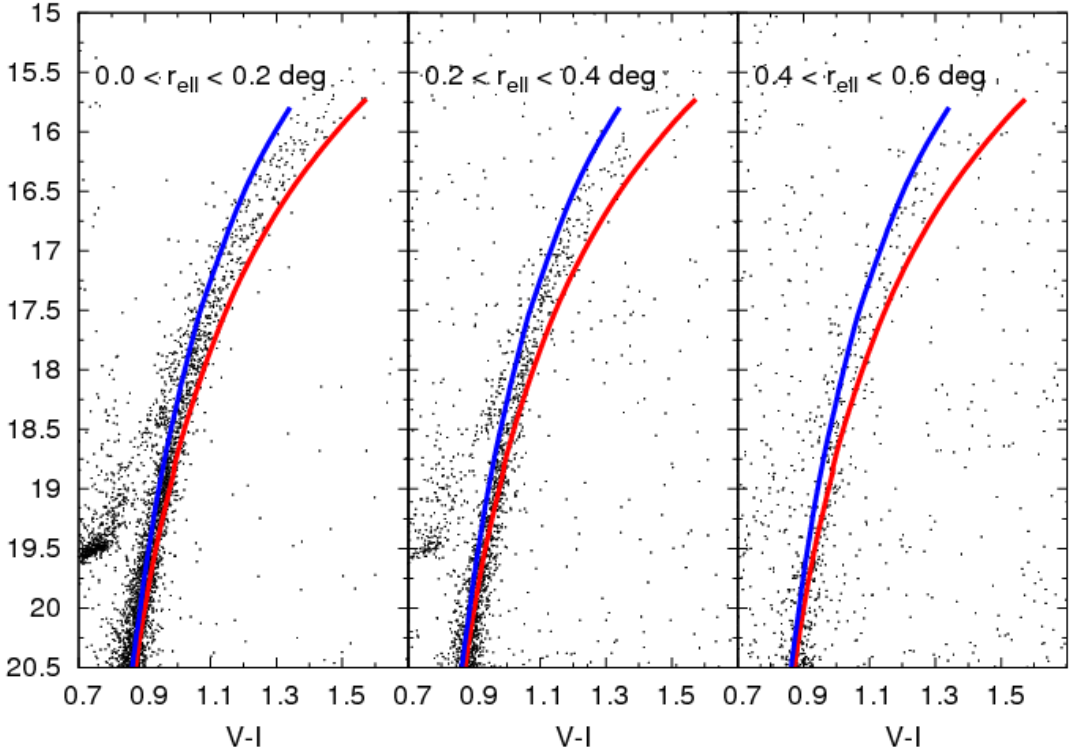}
\caption{The distribution of stars on the Red Giant Branch for different elliptical radii. A metal-rich isochrone~([Fe/H]$\approx-$1.10, [$\alpha$/Fe]$\approx-$0.10, age=7Gyr, red) and metal-poor isochrone~([Fe/H]$\approx-$2.45, [$\alpha$/Fe]$\approx$0.40, age=14Gyr, blue) from the Dartmouth Stellar Evolution Database have been overlaid in each panel. \label{radRGB}} 
\end{figure*}
\\
On the RGB, low resolution \ion{Ca}{ii} triplet (R$\sim$6500) spectroscopy is available for $\approx$630 individual stars using VLT/FLAMES~\citep{Tolstoy04, Battaglia07, Starkenburg10}. We correlated this information with our photometric catalogue to combine the colors, magnitudes and [Fe/H] for a sample of these stars. In Fig.~\ref{RGBCaT}, we show an (I,V$-$I) CMD of the RGB region for different radii overlaid with [Fe/H] metallicities for probable members stars to the Sculptor dSph where \ion{Ca}{ii} spectroscopy is available~\citep{Starkenburg10}. Besides member RGB stars, numerous AGB stars are visible as well, along with a number of possible foreground stars.
\begin{figure*}[!ht]
\centering
\includegraphics[angle=0, width=0.95\textwidth]{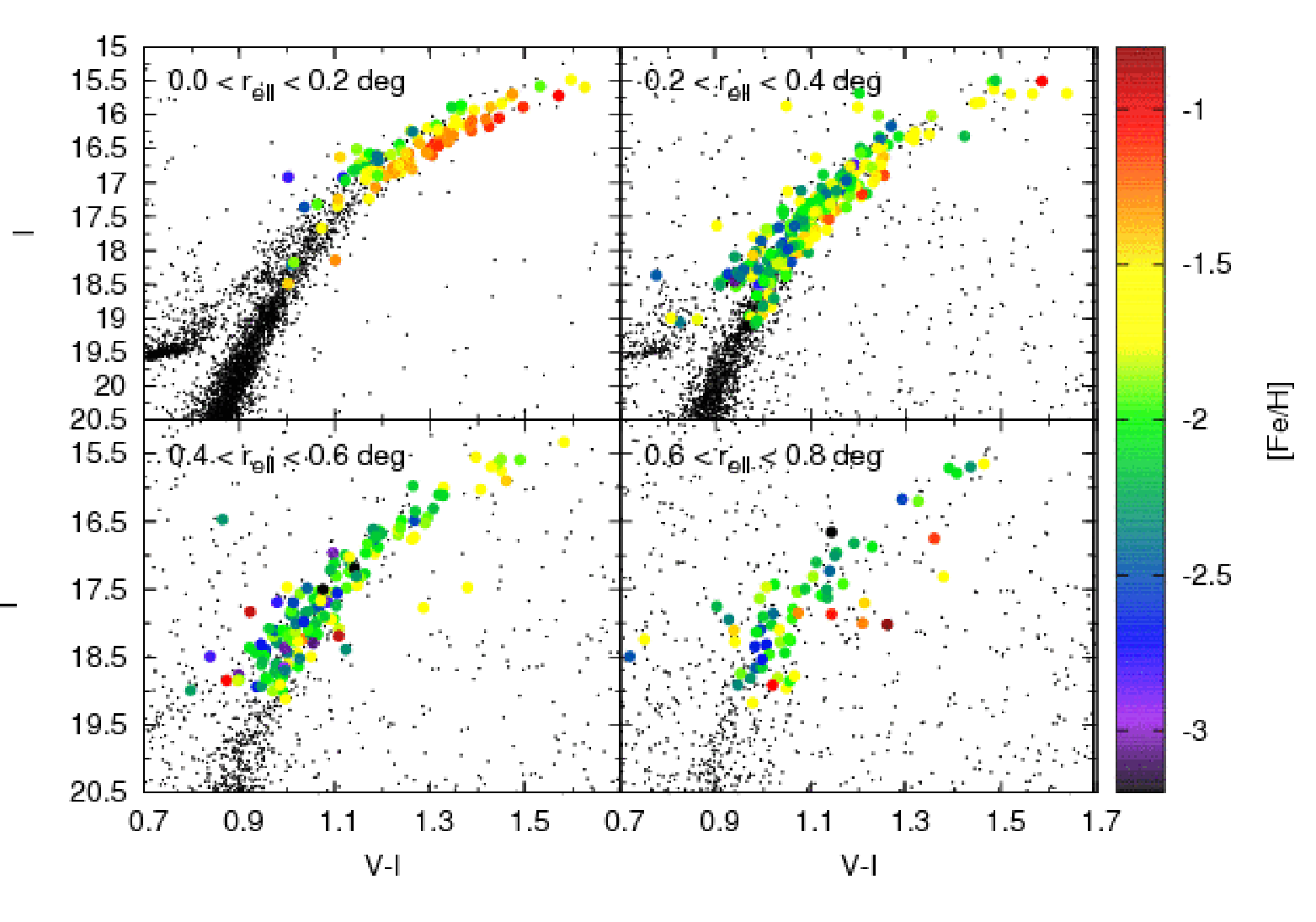}
\caption{Sculptor CMD in the RGB region at different elliptical radii, r$_{ell}$  with Sculptor photometry as (black) dots, overlaid with larger (coloured) filled circles are stars for which spectroscopic metallicities are available. The colours indicate the [Fe/H] metallicities of individual RGB stars~\citep{Starkenburg10}, and the scale is given at the side of each CMD. \label{RGBCaT}} 
\end{figure*}
\\
In the central region low resolution targets were selected only from stars that also had high resolution abundances, which explains the brighter magnitude limit in the top left panel of Fig.~\ref{RGBCaT}. Fig.~\ref{RGBCaT} shows that in the inner region stars are present with metallicities ranging from [Fe/H]=$-$1.0 to [Fe/H]$>-$3.0. As we go outward in the galaxy the more metal rich components which are present on the bright, red part of the RGB in the inner 0.2 deg disappear. This is in agreement with the radial trend observed from the RGB photometry in Fig.~\ref{radRGB}. 
 
\subsection{The Main Sequence Turn-Offs}
\label{int:MSTO}
The MSTO region of the Sculptor dSph has previously only been studied using small fields of view. These studies determined an age range of 13$\pm$2 Gyr~\citep{DaCosta84, Dolphin02}. We present here the first wide-field study of the MSTO population using homogeneous photometry that covers a large fraction of the entire galaxy. \\
A close-up of the MSTO region in the (I,V$-$I) CMD for different elliptical radii is shown in Fig.~\ref{radMS}, overlaid with the same two isochrones used in Fig.~\ref{radRGB}~(one relatively metal poor and old and one metal rich and younger). For increasing distance from the centre we see that the MSTO properties change. The inner regions display both young metal-rich stars and old metal-poor stars, while the outer regions lacks the young, metal-rich population.
\begin{figure*}[!htb]
\centering
\includegraphics[angle=0, width=1.0\textwidth]{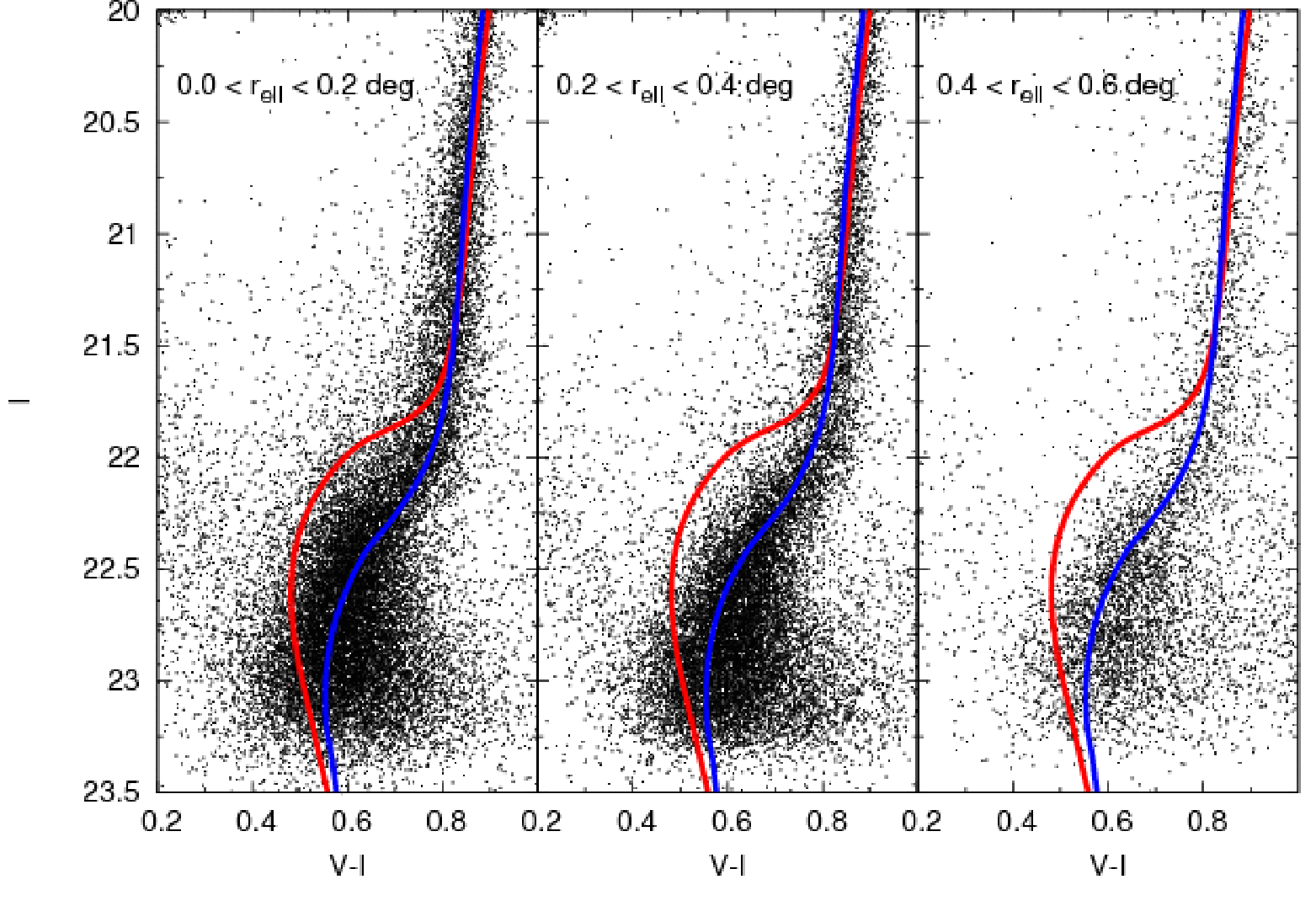}
\caption{The distribution of stars on the Main Sequence Turn-Off for different elliptical radii. Two isochrones representative of a metal-rich~([Fe/H]$\approx-$1.10, [$\alpha$/Fe]$\approx-$0.10, age=7Gyr, red) and a metal-poor~([Fe/H]$\approx-$2.45, [$\alpha$/Fe]$\approx$0.40, age=14Gyr, blue) population are overlaid in each panel. \label{radMS}} 
\end{figure*}
\\
Using isochrones which match the spectroscopic metallicities on the RGB, we have selected regions of the MSTO corresponding to different metallicities and ages~(see Fig.~\ref{MShist}). Regions were defined by selecting stars within a certain distance from each isochrone with the given age and metallicity. The range of [Fe/H] and [$\alpha$/Fe] abundances of the isochrones were chosen to cover the abundance trend derived from HR spectroscopy~\citep[Hill et al., in prep]{Tolstoy09}, with an equal spacing from metal-rich~([Fe/H]=$-$1.00) to metal-poor~([Fe/H]=$-$2.20). The ages of the isochrones were determined by comparing them in the CMD to stars with the same abundance range. Fig.~\ref{MShist} shows the selected regions in the CMD as well as the resulting radial histograms along with the distributions of the RHB and BHB stars. From the MSTO radial distributions we see that the selected regions form a gradient from the metal-poor~(BHB) to the metal-rich~(RHB) distributions. This shows for the first time that this gradient is not only in metallicity, but also in age. The presence of an extended star formation history was expected from studies of the brighter CMD features, but now for the first time the effects of the age gradient can be seen in the radial distribution. The oldest, most metal poor populations~(with [Fe/H]$\approx-$2.5) were formed roughly 14 Gyr ago, while the youngest most metal rich populations~(with [Fe/H]$\approx-$1.0) were formed approximately 7 Gyr ago.

\subsection{Blue Straggler stars}
In the Sculptor CMDs (Fig.~\ref{SclVIrad} \& \ref{SclBVrad}) a BSS population appears to lie above the MSTO region (see labels in Fig.~\ref{SclVIboxes}). In terms of understanding the evolutionary history of Sculptor it is important to understand what process produced these stars. Specifically, are they the result of star formation that continued until the much more recent times, after the main bulk of star formation 7-14~Gyr ago? There is a mechanism of creating BSS stars proposed for globular clusters resulting from mass transfer between low mass stars~\citep{McCrea64}, or collisional processes~\citep{Sigurdsson94}, which relates to an evolution of the old population in Sculptor. 
\begin{figure*}[!thb]
\centering
\includegraphics[angle=0, width=0.92\textwidth]{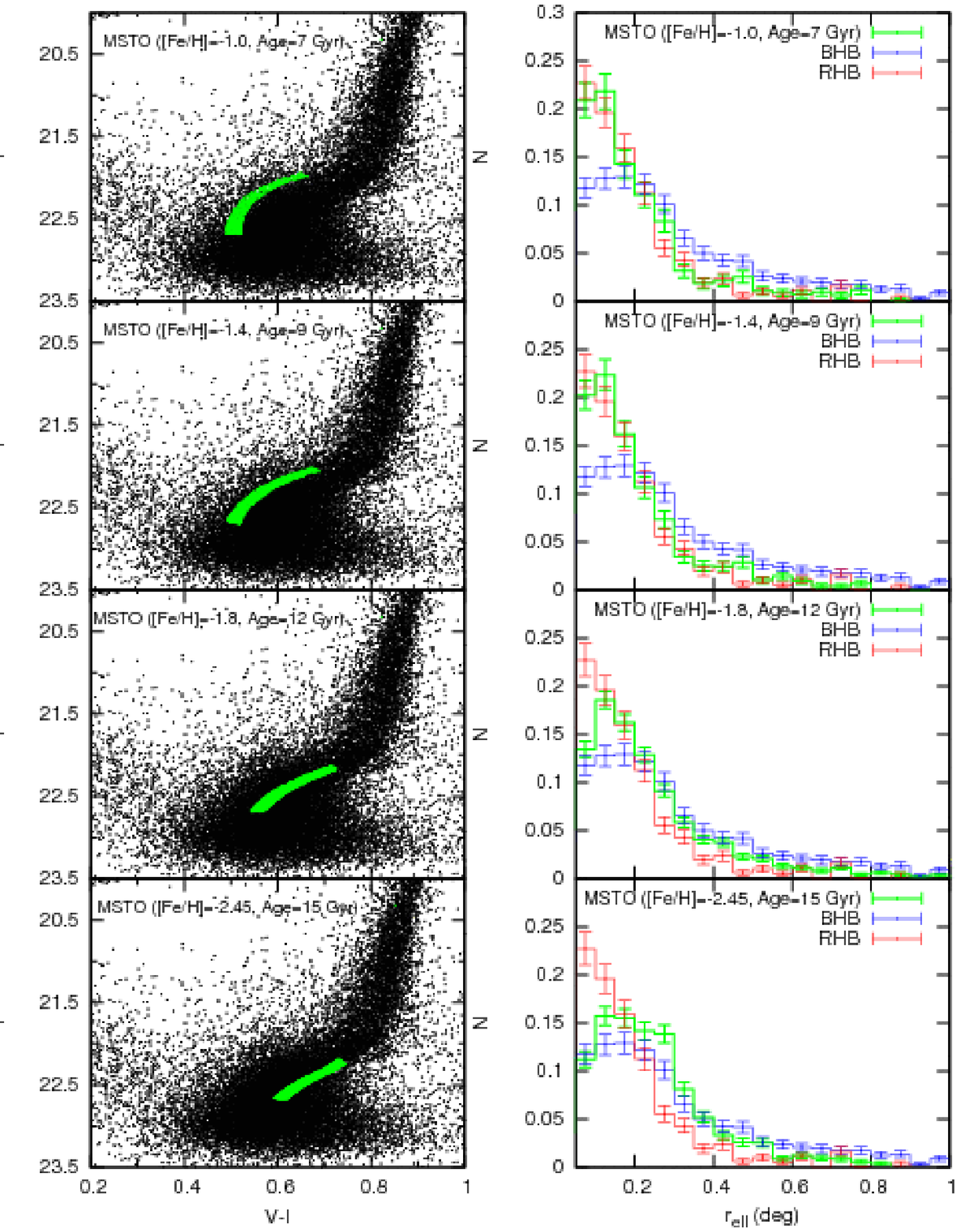}
\caption{On the left are selected regions on the MSTO in green for the labelled age and metallicity. The corresponding radial histograms are shown on the right~(in green). Radial histograms of BHB~(blue) and RHB~(red) stars are also plotted. The radial distributions of the selected regions change from RHB-like for high metallicity to BHB-like for low metallicity. \label{MShist}} 
\end{figure*}
\\
The radial distribution of the BSS population is shown in Fig.~\ref{BSShist}, along with the distributions of RGB and MSTO. The distributions of the other features have been scaled to same total number of stars as the BSS distribution. There is a good agreement between the distributions of BSS and RGB or MSTO, which represent the overall population in Sculptor. This seems to indicate that the BSS population is not linked to any distinct age or metallicity, but likely also has the same age and metallicity gradient as seen in the other CMD features. \\
No obvious trend of a radial gradient in the BSS population is evident in the CMDs with increasing radius, indicating that the BSS population can not easily be separated in different components. In Fig.~\ref{BSShist} there is also no evidence of a central concentration of the BSS population with respect to the overall Sculptor population, and hence no indications of collisional BSS formation as seen in Globular Clusters~\citep{Ferraro97}. This is in agreement with the fact that the central luminosity density of dSph's (0.055 $L_{\odot}/pc^{-3}$ for Sculptor, see~\citet{Mateo98}) is significantly lower than observed in globular clusters (3200 $L_{\odot}/pc^{-3}$ for M3, see~\citet{Harris96}). This means the collision rates in dSph's must be much lower than in globular clusters, which makes the collisional BSS formation channel less likely to be dominant than the binary channel. \\
In order to test if the BSS population is made up of a "normal" young population we modelled the distribution of BSS stars in the CMDs using isochrones, covering the metallicity range observed from spectroscopy, assuming an age of 3 Gyr. Taking completeness effects into account the observed population of $\approx$1600 BSS stars should lead to $\approx$35 stars on the upper RGB. Thus, from the photometry alone it is not possible to rule out that the BSS are a "normal" young population. However, if the BSS stars were a young population they would be expected to be even more centrally concentrated than the older populations, if the trend of increasing central concentration with younger ages holds for these stars. Since this is not observed in Fig.~\ref{BSShist}, BSS formation through mass transfer seems most likely, which results in a distribution similar to that of other populations. This is consistent with results found by~\citet{Mapelli07,Mapelli09} for several nearby dwarf spheroidal galaxies, including Sculptor.

\section{Discussion}
\label{discussion}
We have shown that the two distinct stellar populations found in the Sculptor dSph by~\citet{Tolstoy04} using imaging of HB stars and spectroscopy of RGB stars are linked not only to a metallicity gradient but also to an age gradient going from roughly 7 to 14 Gyr ago. The Main Sequence Turn-Offs~(Figs.~\ref{radMS} \& \ref{MShist}) show unequivocally that the outer regions of Sculptor lack significant numbers of young stars~($<$10 Gyr). This is supported by the spectroscopic \ion{Ca}{ii} triplet abundances~(Fig.~\ref{RGBCaT}), which show a decrease in metal-rich stars with increasing r$_{ell}$. \\
By using a combination of photometry and spectroscopic abundance distributions, we determine the age gradient using the MSTO~(Fig.~\ref{MShist}). This shows that a radial gradient is present in age and metallicity, which is observed also as a bimodality in the HB distribution and as a spread on the RGB and MSTO.  The importance of the MSTO is shown by the fact that only using this part of the CMD it is possible to constrain the age gradient accurately. Linking the age gradient to the existing metallicity gradient suggests that the observed change in HB morphology is due not only to a metallicity gradient~(as suggested by~\citet{Harbeck01}), but that age is also an important parameter. \\
No obvious radial trend with age or metallicity is visible in the BSS population, but there are relatively few stars and there is no obvious way to separate them into distinct components. However, since its radial distribution does match the distribution seen in the overall Sculptor populations~(Fig.~\ref{BSShist}), it is possible that the gradient is also present there. 
\begin{figure}[!ht]
\centering
\includegraphics[angle=270, width=0.45\textwidth]{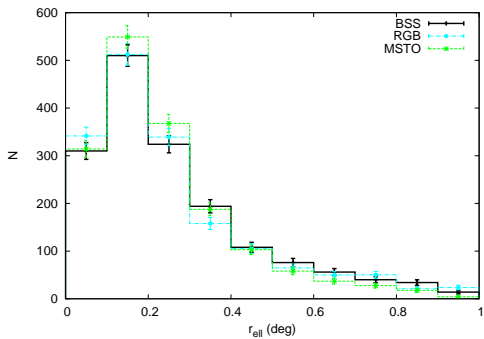}
\caption{Radial distributions of BSS stars~(black crosses) compred to RGB~(light blue filled circles), and MSTO~(green filled squares). \label{BSShist}} 
\end{figure}
\\
These results are consistent with the picture where the Sculptor dSph first formed an extended old metal-poor component, after which later generations of progressively more metal-rich stars were formed ever more concentrated towards the centre of the galaxy. It seems that about 7~Gyr ago the gas supply ran out and the galaxy continued passively evolving up to the present time. \\
The fact that the population gradient is still visible today suggests that it is unlikely that the inner parts of Sculptor have suffered any significant tidal disruption, which could mix the different stellar populations together. Indeed, the spatial distributions~(Fig.~\ref{contours}) do not show any signs of obvious recent tidal interactions. This is in agreement with predictions of theoretical models of the Sculptor dSph~\citep{Sales10}. One aspect which remains unclear is what has caused the increasing ellipticity with radius~\citep{Battaglia07}. This may be due to a very early disruption, which would be visible only in the oldest outer parts of Sculptor, beyond radii studied here. Additionally, it could also be due to an encounter that only affected the outer regions of Sculptor. \\
In a forthcoming paper the CMDs we have presented here will be used to derive a detailed Star Formation History of the Sculptor dSph. This will be combined with high resolution spectroscopic information~(Hill et al., in prep) to derive the enrichment timescales of various elements, and how this fits into the star formation history timescale.

\begin{acknowledgements}
The authors thank ISSI (Bern) for support of the team "Defining the full life-cycle of dwarf galaxy evolution: the Local Universe as a template". T.d.B. would like to thank David Hogg and Dustin Lang for their help in creating a MCMC code to obtaining photometric solutions. T.d.B., E.T. and G.F. gratefully acknowledge the Netherlands Foundation for Scientific Research (NWO) for financial support.
\end{acknowledgements}

\bibliographystyle{aa}
\bibliography{references}

\clearpage
\begin{appendix}
\onecolumn
\section{List of standard field observations}
\label{standardslist}
Observations of standard star fields were carried out during the observing runs in September 2008 and November 2009 in order to provide an accurate calibration of the science observations. Fields were selected from the Landolt standard fields~\citep{Landolt07,Landolt92}, to be used together with Stetson standard star photometry~\citep{Stetson00}. Using the CTIO 4m/Blanco telescope observations of standard star fields were taken during the night in between science observations, in order to monitor the photometric fidelity throughout the night. Standard fields were selected to cover a broad range of airmass and colour, to allow adequate determination of the airmass coefficient and colour term in the photometric solution. In order to ensure of the most accurate calibration possible additional observations of standard stars and Sculptor fields were taken using the CTIO 0.9m telescope during 3 nights in 2008 under photometric conditions. Using these observations the calibration of the 4m Sculptor fields using the 4m standards can be checked and adjusted to the photometric solution obtained under photometric conditions on the 0.9m telescope. Table~\ref{4mstds} lists the observations of standard fields taken with the 4m Blanco telescope, during the runs in 2008 and 2009. The names of the standard fields are given, along with information on exposure times, seeing conditions~(on image) and airmass. Table~\ref{0p9mstds} lists the observations of standard fields taken with the 0.9m telescope, under photometric conditions during the 2008 run. Field names are listed, as well as exposure times, airmass and seeing conditions on image.  
\begin{table*}[!h]
\caption[]{List of observed standard star fields with the 4m CTIO Blanco telescope during two observing runs. Information is given about exposure time, airmass and seeing conditions as determined on image. \label{4mstds}}
\footnotesize
\begin{center}
\begin{tabular}{lcccccc}
\hline\hline
Field		&  RA		& DEC		& Filter & exp time	& seeing    & airmass  \\
		& J2000		& J2000		& 	 & sec		& $^{\prime\prime}$ &        \\	
\hline
 \multicolumn{7}{c}{2008} \\
\hline
T Phe		& 00:30:34.4	& $-$46:28:08	& B	& 10, 8	& 0.9-1.1,1.9	& 1.05,1.08,1.20,2.66 	\\
		&		&		& V	& 5, 4	& 0.8-1.0,1.9 	& 1.05,1.08,1.17,2.62 	\\
		&		&		& I	& 4, 3	& 0.8-1.0,1.8 	& 1.05,1.08,1.17,2.55 	\\
MCT 0401-4017	& 04:03:01.2 	& $-$40:11:28	& B	& 10, 8	& 0.9,1.3	& 1.02, 1.03	   	\\
		&		&		& V	& 5,4	& 1.0-1.1 	& 1.02, 1.03 	   	\\
		&		&		& I	& 4,3	& 1.0-1.1 	& 1.02, 1.03	   	\\
MCT 2019-4339	& 20:22:51.5 	& $-$43:29:24	& B	& 10	& 1.3		& 1.1	           	\\
		&		&		& V	& 5	& 1.4 		& 1.1 		   	\\
		&		&		& I	& 4	& 1.1		& 1.1		   	\\
JL 82		& 21:36:01.0 	& $-$72:47:51	& B	& 10	& 1.6,1.8	& 1.57, 1.91       	\\
		&		&		& V	& 5	& 1.5,1.8	& 1.56, 1.92 	   	\\
		&		&		& I	& 4	& 1.2,1.5 	& 1.55, 1.93	   	\\
\hline
 \multicolumn{7}{c}{2009} \\
\hline
T Phe		& 00:30:34.4	& $-$46:28:08	& B	& 30, 10 & 0.9-1.1	& 1.04,1.05 	\\
		&		&		& V	& 30, 5	 & 0.8-1.0 	& 1.04,1.05 	\\
		&		&		& I	& 30, 4  & 0.7-1.0	& 1.04,1.05 	\\
MCT 0401-4017	& 04:03:01.2 	& $-$40:11:28	& B	& 10	& 0.9,1.7	& 1.42,1.44	   	\\
		&		&		& V	& 5	& 1.0,1.3 	& 1.41,1.44 	   	\\
		&		&		& I	& 4	& 0.9,1.3 	& 1.40,1.45	   	\\
MCT 0550-4911	& 05:51:53.9 	& $-$49:10:31	& B	& 10	& 1.0,1.6	& 1.09,1.73	   	\\
		&		&		& V	& 5	& 0.9,1.2 	& 1.10,1.72 	   	\\
		&		&		& I	& 4	& 1.0,0.9 	& 1.10,1.70	   	\\
WD 0830-535	& 08:31:56.5 	& $-$53:40:52	& B	& 10	& 1.5		& 1.13	   	\\
		&		&		& V	& 5	& 1.1	 	& 1.13 	   	\\
		&		&		& I	& 4	& 0.9	 	& 1.12	   	\\
LB 1735		& 04:31:34.7 	& $-$53:37:13	& B	& 10	& 1.0		& 1.1	   	\\
		&		&		& V	& 5	& 0.9	 	& 1.1 	   	\\
		&		&		& I	& 4	& 0.8	 	& 1.1	   	\\
L92		& 00:55:38.0 	& +00:56:00.0	& B	& 30, 5	& 1.5		& 1.18	   	\\
		&		&		& V	& 30, 5	& 1.2	 	& 1.17 	   	\\
		&		&		& I	& 30, 5	& 1.3	 	& 1.17	   	\\
L101		& 09:56:39.0 	& $-$00:19:53.0	& B	& 30, 5	& 1.8		& 1.42	   	\\
		&		&		& V	& 30, 5	& 1.6	 	& 1.45 	   	\\
		&		&		& I	& 30, 5	& 1.2	 	& 1.49	   	\\
PG 2213-006	& 22:16:33.5 	& $-$00:23:34.7	& B	& 30, 5	& 1.5		& 1.27	   	\\
		&		&		& V	& 30, 5	& 1.2	 	& 1.29 	   	\\
		&		&		& I	& 30, 5	& 1.2	 	& 1.30	   	\\
PG 0231+051	& 02:33:55.3 	& +05:14:22.9	& B	& 30, 5	& 1.3		& 1.60	   	\\
		&		&		& V	& 30, 5	& 1.1	 	& 1.57 	   	\\
		&		&		& I	& 30, 5	& 1.0	 	& 1.54	   	\\
L95		& 03:53:58.0 	& +00:03:30.0	& B	& 30, 5	& 1.5		& 2.10	   	\\
		&		&		& V	& 30, 5	& 1.4	 	& 2.00 	   	\\
		&		&		& I	& 30, 5	& 1.4	 	& 1.90	   	\\
L95		& 06:51:58.7 	& $-$00:24:18.7	& B	& 30, 5	& 1.0		& 1.17	   	\\
		&		&		& V	& 30, 5	& 1.0	 	& 1.17 	   	\\
		&		&		& I	& 30, 5	& 0.9	 	& 1.16	   	\\
\hline
\hline 
\end{tabular}
\end{center}
\end{table*}

\clearpage

\begin{table*}[!ht]
\caption[]{List of observed standard star fields with the 0.9m CTIO telescope. Information is given about exposure time, airmass and seeing conditions as determined on image. \label{0p9mstds}}
\footnotesize
\begin{center}
\begin{tabular}{lcccccc}
\hline\hline
Field		&  RA			& DEC		& Filter	& exp time	& seeing		& airmass \\
		& J2000			& J2000		& 		& sec		& $^{\prime\prime}$ 	&    \\	
\hline
JL 163		& 00:10:42.6 	& $-$50:14:10	& B	& 150	& 1.4, 1.8-2.0	& 1.16,2.31	\\
		&			&			& V	& 150	& 1.3, 1.8-2.1	& 1.15, 2.39	\\
		&			&			& I	& 300	& 1.2, 1.6-1.7 	& 1.13, 2.53	\\
T Phe		& 00:30:34.4 	& $-$46:28:08	& B	& 150	& 1.4-1.7, 2.2-2.4 & 1.04-1.12, 1.21-1.26, 1.57 \\
		&			&			& V	& 150	& 1.3-1.7, 2.0 	& 1.04-1.13, 1.20-1.28, 1.60 	\\
		&			&			& I	& 300	& 1.3-1.7, 2.2-2.4 	& 1.04-1.15, 1.18-1.34, 1.65	\\
MCT 0401-4017	& 04:03:01.2 & $-$40:11:28	& B	& 150	& 1.4-1.5, 2.1	& 1.04-1.07, 1.16-1.24, 2.16	\\
		&			&			& V	& 150	& 1.3-1.6, 2.4 	& 1.05-1.08, 1.18-1.20, 2.08 	\\
		&			&			& I	& 300	& 1.2-1.5, 2.0 	& 1.04-1.09, 1.21-1.18, 2.00	\\
LB 1735		& 04:31:34.7 & $-$53:37:13	& B	& 150	& 1.3, 2.3	& 1.11-1.17, 1.38 \\
		&			&			& V	& 150	& 1.4, 1.9 	& 1.11-1.18, 1.37 \\
		&			&			& I	& 300	& 1.2, 2.0	& 1.12-1.20, 1.34 \\
LSE 259		& 16:53:45.1 & $-$56:00:16	& B	& 150	& 4.0		& 2.16	\\
		&			&			& V	& 150	& 2.2 		& 2.21 	\\
		&			&			& I	& 300	& 2.3		& 2.29	\\
JL 82		& 21:36:01.0 & $-$72:47:51	& B	& 150	& 1.8		& 1.38-1.40 \\
		&			&			& V	& 150	& 1.5-1.7	& 1.39-1.41 \\
		&			&			& I	& 300	& 1.4		& 1.40-1.42 \\
JL 117		& 22:54:55.5 & $-$72:22:59	& B	& 150	& 1.5		& 1.35	\\
		&			&			& V	& 150	& 1.4 		& 1.35 	\\
		&			&			& I	& 300	& 2.0		& 1.35	\\
\hline
\hline 
\end{tabular}
\end{center}
\end{table*}
\clearpage

\end{appendix}

\begin{appendix}
\onecolumn
\section{Photometric calibration}
\label{calappendix}
Obtaining an accurate photometric calibation is essential for a reliable interpretation of a CMD. For example the absolute luminosity
of a Main Sequence Turn-off star correlates with age. The observing strategy was chosen in such a way as to combine deep 4m science images with 0.9m calibration images taken under photometric conditions. These images are used to ensure the most accurate photometric calibration possible for the science images. \\
A comparison of the calibrated magnitude and true magnitude of the standard stars used to obtain the photometric solution versus observed magnitude~(for both the 4m and 0.9m data) is shown in Fig.~\ref{magdiff}. Similar plots showing the magnitude difference versus airmass and colour are shown in Figs.~\ref{airdiff} and~\ref{coldiff} respectively. A line indicating the zero level residual~(black) and the mean of the residual~(blue) is shown, along with errorbars denoting the average photometric error. These plots show the accuracy of the calibration applied to the science data. The standard deviation of the residuals is $\sigma_{B}\approx$0.033, $\sigma_{V}\approx$0.028, $\sigma_{I}\approx$0.032 for the 4m data and $\sigma_{B}\approx$0.038, $\sigma_{V}\approx$0.028, $\sigma_{I}\approx$0.029 for the 0.9m data. Fig.~\ref{magdiff} shows that the zeropoint correction is consistent across all magnitudes used in the photometric solution, with the mean level of the residuals~(blue line) being less than the average photometric error. Figs.~\ref{airdiff} and~\ref{coldiff} show that the effects of colour and airmass are well taken care of, with the mean level of the residuals~(blue line) being less than the average photometric error. Residuals in the B band observations of the 0.9m telescope are most offset from zero, due to the poorer seeing conditions and larger photometric errors in the B band. However, the B band residuals are consistent with zero in all plots, given the average photometric error in B, showing that an accurate calibration was obtained.
\begin{figure*}[!h]
\centering
\includegraphics[angle=270, width=0.85\textwidth]{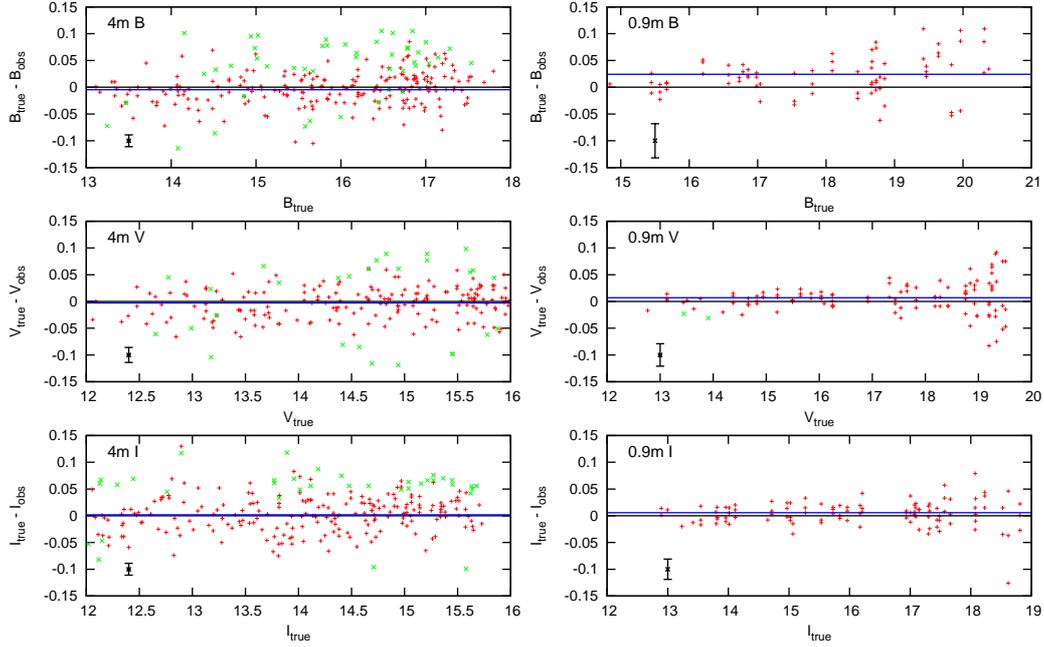}
\caption{Comparison between calibrated and true magnitudes for the standard stars used to obtain the photometric solution. Each plot shows magnitude residuals vs. true magnitude for different filters for the 4m data~(left) and 0.9m data~(right). Stars classified as outlyers when determining the photometric solution are marked in green. The average standard deviation of the residuals over all filters is $\approx$0.033 for the 4m data, and $\approx$0.03 for the 0.9m data. A line indicating the zero level residual~(black) and the mean of the residual~(blue) is also shown~(visible only if there is an offset from zero), along with errorbars denoting the average photometric error. The zeropoint correction residuals are consistent with zero across all magnitudes, given the average photometric error. The mean of the residual in the 0.9m B~(upper right) solution is most offset from zero~(by 0.024) due to seeing conditions and large photometric error in B. However, the residuals are still within the average photometric error~($\pm$0.032), indicating an accurate calibration is achieved. \label{magdiff}} 
\end{figure*}
\begin{figure*}[!ht]
\centering
\includegraphics[angle=270, width=0.85\textwidth]{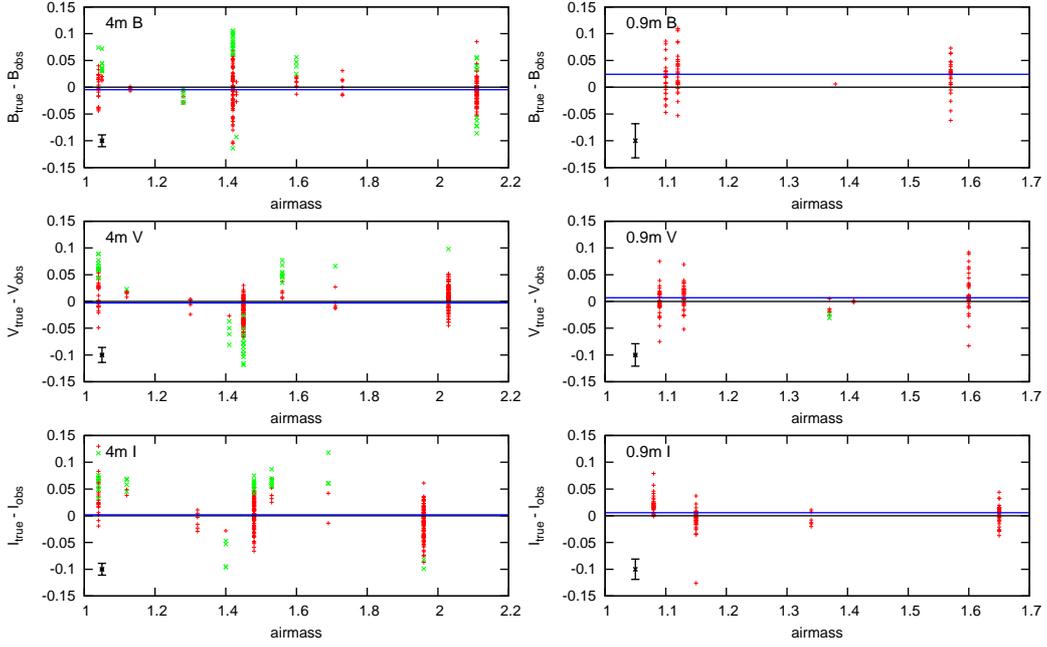}
\caption{Magnitude residuals vs. airmass for photometric standard stars for the 4m data~(left) and 0.9m data~(right). See Fig.~\ref{magdiff} for details. No trend with increasing airmass is visible, showing that the effects of airmass are well taken care of in the photometric calibration. \label{airdiff}} 
\end{figure*}
\begin{figure*}[!ht]
\centering
\includegraphics[angle=270, width=0.85\textwidth]{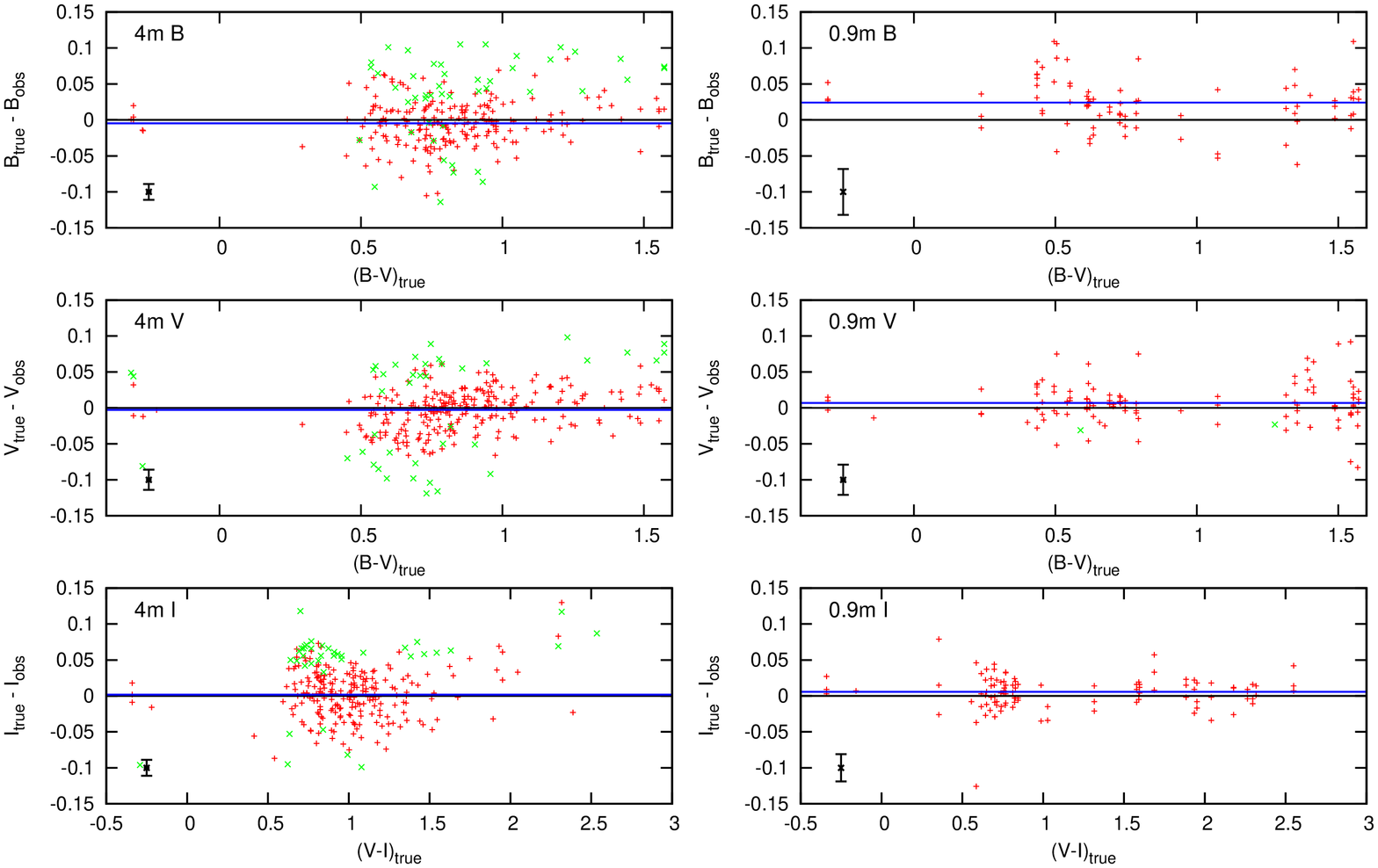}
\caption{Magnitude residuals vs. colour for 4m~(left) and 0.9m~(right).standard stars. See Fig.~\ref{magdiff} for details. No trend with colour is visible, showing that the photometric calibration correctly takes care of the effects of the colour term. \label{coldiff}} 
\end{figure*}
\\
A second check of the photometry is done by comparing aperture and PSF magnitudes.We determined aperture magnitudes for small samples of stars in standard fields and in the central Sculptor pointing. A photometric solution was obtained from the aperture photometry, and used to calibrate the aperture magnitudes in the Sculptor central field. Then, a comparison of the PSF and aperture magnitudes is made~(shown in Fig.~\ref{apdopdiff}) for both the 0.9m and 4m data. The standard deviation of the residuals is $\sigma_{B}\approx$0.101, $\sigma_{V}\approx$0.040, $\sigma_{I}\approx$0.033 for the 4m data and $\sigma_{B}\approx$0.108, $\sigma_{V}\approx$0.028, $\sigma_{I}\approx$0.029 for the 0.9m data. Fig.~\ref{apdopdiff} also shows that the differences between aperture and PSF magnitudes are negligible within the average photometric errors, giving confidence to the reliability of our PSF photometry. The B band residual is once again most offset from zero~(by 0.05), but consistent with zero given the photometric errors~($\pm$0.061). \\
Next, the calibrated 4m and 0.9m PSF mags and aperture mags are compared to check whether the 4m data is consistently calibrated with the 0.9m data taken under photometric conditions. A comparison of both data sets for the central Sculptor field is shown in Fig.~\ref{4m0p9diff} for PSF and aperture magnitudes respectively. Solid black errorbars have been overplotted for the PSF magnitudes, indicating the average error on the magnitude difference at bright, intermediate and faint magnitudes. For all three selections the mean of the residual is smaller than the error on the magnitude difference~(from bright to faint: \textbf{B:} 0.0043$<$0.0617,0.0079$<$0.0880,0.0095$<$0.1546, \textbf{V:} 0.0099$<$0.0176,0.0154$<$0.0484,0.0187$<$0.0890, \textbf{I:} 0.0002$<$0.0125,0.0072$<$0.0272,0.0185$<$0.0844). For the aperture magnitudes the mean of the residuals~(\textbf{B:} 0.0170 \textbf{V:} 0.0321 \textbf{I:} 0.0236) is comparable to the average error~(\textbf{B:} 0.0759 \textbf{V:} 0.0292 \textbf{I:} 0.0191). The relatively high residuals are possibly due to the low number of stars with aperture magnitudes used in the determination of the photometric solution for the 0.9m data. \\
The above figures show that the 4m calibration properly takes into account zeropoint, airmass and colour effects. Furthermore, the calibration is consistent between aperture and PSF magnitudes, given the photometric errors. The absolute calibration of the 4m data is consistent with that of the 0.9m data for PSF magnitudes and to a lesser extent for aperture magnitudes. Thus, an accurate absolute calibration is achieved for the PSF magnitudes used in the final photometry catalog. The final catalog has an average accuracy due to random errors for all filters of $\approx \pm$0.002 for the brightest stars, while at the faint end the accuracy is $\approx \pm$0.2. The accuracy of the calibration varies for different magnitudes and filters, but the comparison of calibrated and true magnitudes shows an accuracy of~$\approx$0.04 mag or better across the magnitude range used. 
\begin{figure*}[!ht]
\centering
\includegraphics[angle=270, width=0.85\textwidth]{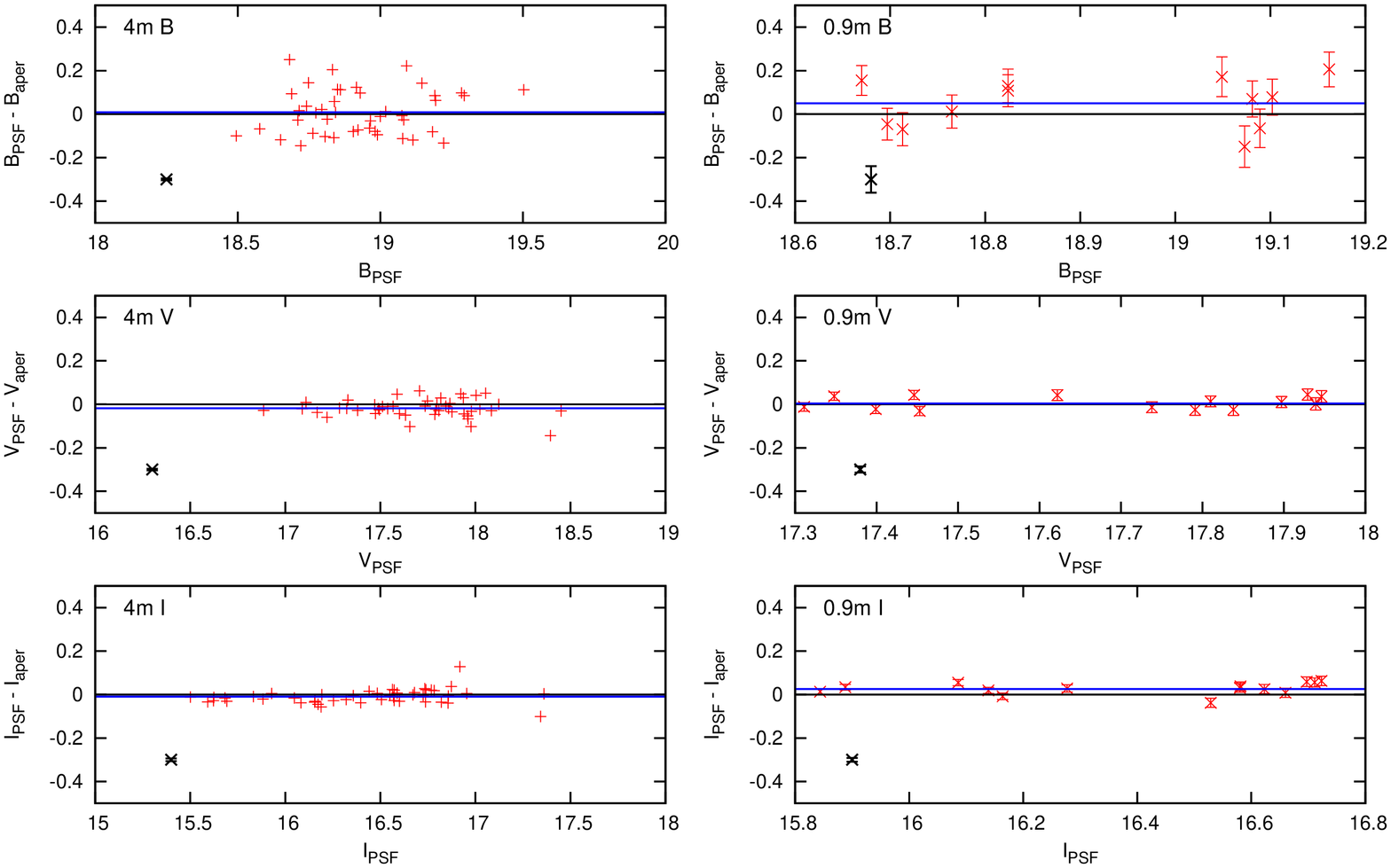}
\caption{Comparison of aperture and PSF magnitudes in the central Sculptor pointing for the 4m data~(left) and 0.9m data~(right).  See Fig.~\ref{magdiff} for details. The residuals are consistent with zero, given the photometric errors, showing that the calibrations using aperture and PSF magnitudes agree. \label{apdopdiff}} 
\end{figure*}
\begin{figure*}[!ht]
\centering
\includegraphics[angle=270, width=0.85\textwidth]{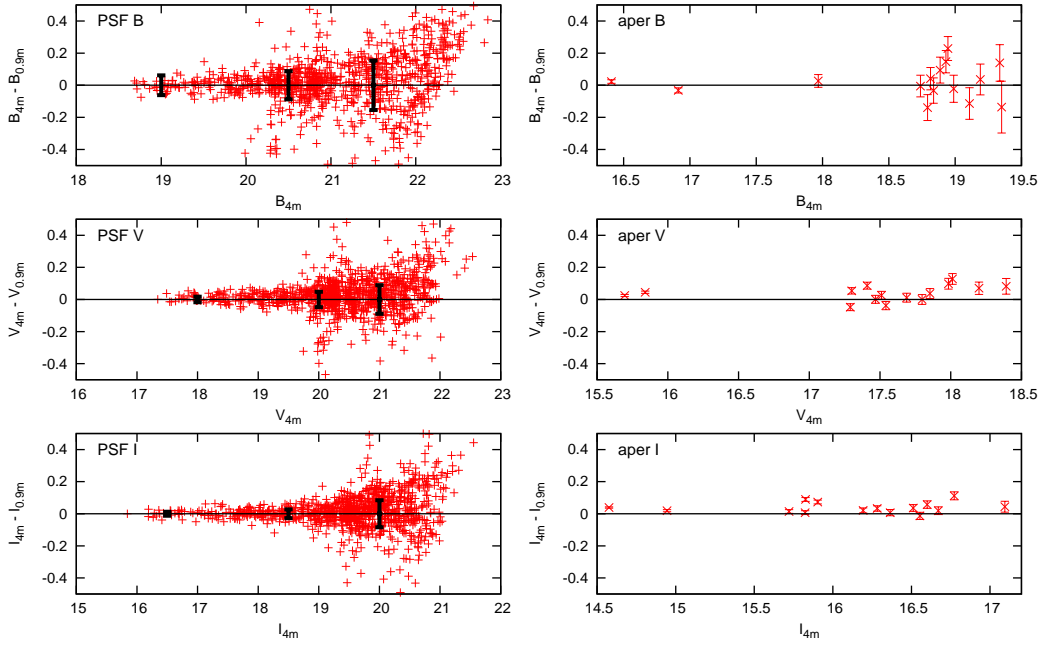}
\caption{Comparison between 4m and 0.9m photometry for the central Sculptor pointing for PSF~(left) and aperture~(right) magnitudes. A plume of short period RR Lyrae variable stars is visible in all bands in the PSF data at B$\sim20.5-21$; V$\sim20-20.5$; I$\sim19.5-20.5$. Solid black errorbars overplotted in the left-hand figures indicate the average error on the magnitude difference at the corresponding magnitude. The residuals of the PSF magnitudes are consistently smaller than the error on the magnitude difference, while for the aperture magnitudes the residuals are similar to the error. \label{4m0p9diff}} 
\end{figure*}

\end{appendix}

\end{document}